\documentclass[twocolumn,prd,floatfix,preprintnumbers,nofootinbib,superscriptaddress]{revtex4-2}
\usepackage{graphicx}
\usepackage{epsfig,float}
\usepackage{dcolumn}
\usepackage{bm}
\usepackage{bbm}
\usepackage{morefloats}
\usepackage{color}
\usepackage{amsmath}
\usepackage{mathrsfs}
\usepackage{slashed}
\usepackage{multirow}
\usepackage{amssymb}
\usepackage{bbding}
\usepackage{hyperref}
\hypersetup{colorlinks=true,linkcolor=blue,citecolor=blue,menucolor=black,urlcolor=black,filecolor=blue}
\usepackage{appendix}
\usepackage{multirow}
\usepackage{graphicx}
\usepackage{subfigure}
\usepackage{subfloat}    
\usepackage{float}
\usepackage{amssymb}
\usepackage{tabularx}
\graphicspath{{./image}}

\makeatletter
\newcommand*{\rom}[1]{\expandafter\@slowromancap\romannumeral #1@}
\makeatother
\newcommand{\bs}[1]{\boldsymbol #1}

\begin{document}

\title{Singly Cabibbo-suppressed hadronic weak decays of the $\Omega^-$ hyperon}

\author{Ye Cao}~\thanks{caoye@impcas.ac.cn}
\affiliation{Southern Center for Nuclear-Science Theory (SCNT), Institute of Modern Physics, Chinese Academy of Sciences, Huizhou 516000, China}

\author{Tao Zhong}~\thanks{zhongtao@gzmu.edu.cn}
\affiliation{School of Physics and Mechatronic Engineering, Guizhou Minzu University, Guiyang 550025, China}

\author{Ju-Jun Xie}~\thanks{xiejujun@impcas.ac.cn (corresponding author)}
\affiliation{Southern Center for Nuclear-Science Theory (SCNT), Institute of Modern Physics, Chinese Academy of Sciences, Huizhou 516000, China}
\affiliation{Heavy Ion Science and Technology Key Laboratory, Institute of Modern Physics,
Chinese Academy of Sciences, Lanzhou 730000, China}
\affiliation{School of Nuclear Sciences and Technology, University of Chinese Academy of Sciences, Beijing 101408, China}

\author{Qiang Zhao}~\thanks{zhaoq@ihep.ac.cn}
\affiliation{Institute of High Energy Physics, Chinese Academy of Sciences, Beijing 100049, China}
\affiliation{University of Chinese Academy of Sciences, Beijing 100049, China}
\affiliation{Center for High Energy Physics, Henan Academy of Sciences, Zhengzhou 450046, China}

\begin{abstract}

We study the two-body hadronic weak decays of the $\Omega^-$ hyperon with strangeness $S=-3$, including three singly Cabibbo-suppressed decay modes: $\Xi^0 \pi^-$, $\Xi^-\pi^0$ and $\Lambda K^-$. The decay amplitudes at the quark level, arising from $s\to ud \bar{u}$ transitions (direct pion emission and color-suppressed processes) and $su\to ud$ transitions (pole terms), are calculated in the framework of the non-relativistic constituent quark model. The theoretical results show that the $\Xi^0 \pi^-$ channel is dominated by the color-allowed direct pion emission process, while the $\Lambda K^-$ channel is well described by one type of pole contribution mediated through intermediate $\Xi$ resonances ($1^2S_{1/2^+}$ and $1^2P_{1/2^-}$ states). However, the contribution from tree-level mechanisms alone to the branching ratio of $\Omega^- \to \Xi^- \pi^0$ is small due to its color-suppressed nature. The discrepancy is resolved by including final state interactions through rescattering processes via intermediate states $\Xi^0\pi^-$ and $\Lambda K^-$. This work demonstrates that a unified description of $\Omega^-$ hadronic weak decays necessitates the interplay of quark-level weak vertices, baryon pole structures, and long-distance final state rescattering dynamics. With these mechanisms, the obtained branching ratios are in agreement with the high-precision experimental data from the BESIII. Furthermore, these above decays are found to be dominated by the parity-conserving $P$-wave transitions, thus the asymmetry parameters are almost zero.

\end{abstract}

\maketitle

\section{Introduction}

Although the $\Omega^-$ hyperon with strangeness $S=-3$ was discovered 60 years ago~\cite{Barnes:1964pd}, its spin was not experimentally determined until 2006~\cite{BaBar:2006omx}. And many of its decay modes are still poorly known experimentally~\cite{ParticleDataGroup:2024cfk}. The $\Omega^-$ occupies a special role in the lowest-lying spin-3/2 decuplet of flavor SU(3) symmetry. All other decuplet states are dominated by the strong decay modes, however, the $\Omega^-$ decays only via the weak interaction. Its primary decay channels, as measured with high precision by the BESIII collaboration, are $\Omega^- \to \Lambda K^-$ with a branching ratio of $(66.3 \pm 0.8 \pm 2.0)\%$, and decays involving a $\Xi$ hyperon, such as $\Omega^- \to \Xi^0 \pi^-$ with a branching ratio of $(25.03 \pm 0.44 \pm 0.53)\%$, and $\Omega^- \to \Xi^- \pi^0$ of $(8.43 \pm 0.52 \pm 0.28)\%$~\cite{BESIII:2023ldd}. Moreover, the first model-independent determination of the $\Omega^-$ spin was conducted to be 3/2 by the BESIII collaboration in 2021~\cite{BESIII:2020lkm} since its discovery more than 50 years, and the decay parameters of the processes $\Omega^-\to\Lambda K^-$ and $\bar{\Omega}^+\to \bar{\Lambda} K^+$ were measured for the first time by a fit to the angular distribution of the complete decay chain. 

On theoretical side, the nonleptonic decay amplitudes of $\Omega^-$ hyperon have been predominantly calculated within the framework of chiral perturbation theory (ChPT), the effective field theory of quantum chromodynamics (QCD) at low energies. A systematic analysis was done in Ref.~\cite{Carone:1991ni}, where baryons were treated as heavy particles in a low-energy chiral Lagrangian. It was found that the three dominant $\Omega^-$ decay modes proceed via $P$-wave and are governed by a single pole diagram at leading order~\cite{Carone:1991ni}. The heavy baryon ChPT was extended to one-loop level in Ref.~\cite{Egolf:1998vj}, incorporating the leading nonanalytic corrections. The role of intermediate resonances was further examined in Ref.~\cite{Borasoy:1999ip}, suggesting that $J^P = {1}/{2}^{\pm}$ resonances play an essential role in these $\Omega^-$ hadronic weak decays. In addition, Ref.~\cite{Duplancic:2001zr} provided theoretical calculations of the branching ratios by means of the QCD enhanced effective weak Hamiltonian supplemented by the SU(3) Skyrme model. Earlier work by Ref.~\cite{Galic:1979hh} calculated the branching ratios using the Weinberg-Salam model with QCD corrections and the MIT bag model. The crucial role played by the non-factorizable contributions is supported by studies with the topological diagram approach (TDA) under SU(3) flavor symmetry~\cite{Xu:2020jfr}. On the other hand, the asymmetry parameter has long been considered to be related to $CP$ violation (CPV), and hyperon decays provide an ideal environment for testing CPV. In Refs.~\cite{Tandean:1998wr,Tandean:2004mv}, the CPV asymmetry in the $\Omega^-\to \Xi\pi$ and $\Omega^-\to \Lambda K^-$ channels was evaluated, and it was found that the partial-rate asymmetry in $\Omega^-\to \Lambda K^-$ is non-vanishing due to final-state interactions (FSIs). The above theoretical efforts have established a tentative and somewhat rough but still evolving picture of $\Omega^-$ decays, where precise predictions for all decay parameters remain challenging and highlight the critical need for precise experimental constraints.

The main purpose of this work is to systematically investigate the hadronic weak decays of $\Omega^-\to \Xi^0\pi^-$, $\Xi^-\pi^0$ and $\Lambda K^-$ within a phenomenological framework. This analysis is motivated by the availability of recent high-precision experimental data~\cite{BESIII:2023ldd}. We employ the nonrelativistic constituent quark model (NRCQM) that has been previously applied to describe the hadronic weak decays of the octet hyperons, such as $\Lambda$ and $\Sigma^\pm$~\cite{Cao:2025kvs}. In these hadronic weak decays, the two-body final states can carry orbital angular momentum $L=1$ ($P$-wave) for parity-conserving (PC) transition or $L=2$ ($D$-wave) for parity-violating (PV) transition. In addition, we consider also the final state interactions mechanism, which is important  in some hyperon hadronic weak decays~\cite{Cheng:2004ru, Yu:2017zst, Cheng:2021nal, Cao:2023csx, Cao:2023gfv, Cao:2025kvs}. By calculating the branching ratios and the asymmetry parameters for the $\Omega^-$ decays, we aim to provide predictions that can be directly compared with existing and forthcoming measurements. We anticipate that the results of this study will contribute significantly to a deeper understanding of the decay dynamics and polarization properties of spin-3/2 hyperons, offering insights into the weak interaction in the strange baryon sector.

As follows, we will first introduce the quark-level mechanism including $s\to ud\bar{u}$ and $su\to ud$ processes and hardon-level final state interaction mechanism in Sec.~\ref{sec-quark-level amp} and Sec.~\ref{sec-hardon-level amp}, respectively. Then, numerical results for the partial widths and asymmetry parameters and discussions will be presented in Sec.~\ref{sec-numerical-results}. Finally, a brief summary is given in Sec.~\ref{sec-summary}.

\section{The quark-level amplitudes and physical observables}\label{sec-quark-level amp}

\begin{figure}[htbp]
    \centering
    \subfigure[\ DPE]{
        \includegraphics[width=0.22\textwidth]{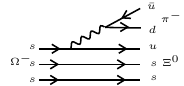}
        \label{fig:sub1}
    }
    \subfigure[\ CS-1]{
        \includegraphics[width=0.22\textwidth]{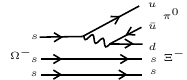}
        \label{fig:sub2}
    }
    \subfigure[\ CS-2]{
        \includegraphics[width=0.22\textwidth]{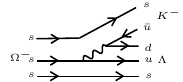}
        \label{fig:sub3}
    }
    \subfigure[\ PT]{
        \includegraphics[width=0.22\textwidth]{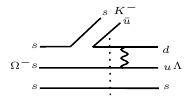}
        \label{fig:sub4}
    }
    \caption{Illustrations for the two-body hadronic weak decay of $\Omega^-$ into $\Xi\pi$ and $\Lambda K^-$ at the quark level. (a) Direct pion emission (DPE) processes; (b)-(c) Color suppressed (CS) pion emission processes; (d) Pole terms (PT).}
    \label{fig:quark-level diagram}
\end{figure}

The corresponding transition diagrams at the quark level for these $\Omega^-\to \Xi^0\pi^-$, $\Xi^-\pi^0$, and $\Lambda K^-$ decays are shown in Fig.~\ref{fig:quark-level diagram}. The $\Xi^0\pi^-$ and $\Xi^-\pi^0$ channels proceed only via the direct pion emission (DPE) process and the color suppressed (CS) process respectively, while the $\Lambda K^-$ channel involves both the color suppressed (CS) process and the pole term (PT) process. The DPE and CS processes share a common quark-level transition mechanism of $s\to ud\bar{u}$. They are distinguished by whether the $W$ boson is emitted externally or internally, thereby a color suppression factor will be present in the CS amplitudes. With the $s$ quark in the initial-state baryon converting to a $u$ quark by emitting a $W^-$ boson, the quark pair $d\bar{u}$ coupled to the $W^-$ will directly hadronize into the final-state meson $\pi^-$ for DPE process. In contrast, the CS process describes the hadronization of the final-state meson with a quark in the initial-state baryon combined with the antiquark created by the $W$ emission. Depending on how the antiquark from the $W$ emission is combined with one quark from the initial baryon, the CS process can be categorized into two types: CS-1 and CS-2, as labeled in Fig.~\ref{fig:quark-level diagram} for $\Omega^-$ decays. PT processes are driven by the quark-level $su\to ud$ internal conversion, with the final-state meson emitted via strong interaction vertices, as illustrated in Fig.~\ref{fig:quark-level diagram} (d). In such a transition the weak and strong interactions occur non-locally, and pole structures in the transition amplitudes can be identified. Typically, pole terms are two-step processes, where the weak transition either precedes or follows the strong meson emission~\cite{Cao:2025kvs}. In this work, however, we find that only one sequence contributes to the $\Lambda K^-$ channel: the weak interaction occurs first, followed by the strong interaction. This indicates that the pole-term mechanism in $\Omega^-$ decays may differ from those described in Refs.~\cite{Cao:2025kvs, Richard:2016hac}.

At low energies, constituent quark degrees of freedom dominate, motivating us to employ the nonrelativistic constituent quark model for the baryon-state description entering transition amplitudes via wave-function convolution. Quark-level operators together with baryon wave functions further expose the underlying dynamical features of the relevant interactions. The spin and flavor wave functions are trivial and listed in Appendix~\ref{app:spin and flavor wf}, while the spatial wave functions can be obtained by numerically solving the Schr\"{o}dinger equation for the baryon state, as presented in Sec.~\ref{sec-baryon wave functions}. Importantly, different transition amplitudes can be connected to each other through the SU(3) flavor symmetry. An interesting and non-negligible feature is that the SU(3) flavor symmetry should be broken and the breaking effects can be accounted for via both constituent quark mass and the wave function convolutions. This feature is different from hadron-level parametrization approaches, where the SU(3) flavor symmetry effects can be absorbed into the coupling constants, leaving the connections and relative phases among different transition amplitudes actually unclear. Besides, within the NRCQM framework, transition amplitudes for both DPE and CS mechanisms can be computed explicitly with no extra free parameters, as these processes are governed by the identical quark-level $s\to ud\bar{u}$ operator. Next, we mainly introduce how to obtain the transition amplitude shown in Fig.~\ref{fig:quark-level diagram} within the theoretical framework of NRCQM.

\subsection{The decay amplitudes of $s\to ud\bar{u}$ processes}

It is generally assumed that the hadronic weak decay are described with a current-current type Hamiltonian. In the nonrelativistic limit, the effective weak Hamiltonian $\hat{H}_{W,1\to 3}^{(\text{PC})}$ and $\hat{H}_{W,1\to 3}^{(\text{PV})}$ for the $s_3\to u_{3^{\prime}}d_5\bar{u}_4$ processes are~\cite{Richard:2016hac,Niu:2020gjw,Li:2025noz}:
\begin{widetext}
    \begin{eqnarray}
\hat{H}_{W,1\to 3}^{(\text{PC})} &=& \frac{G_F}{\sqrt{2}}\frac{\beta V_{ud}V_{us}}{(2\pi)^3}\hat{\alpha}_3^{(+)} \hat{I}_{P} \frac{\delta_{c_4c_5}\delta_{c_3c_3^{\prime}}}{\sqrt{3}} \delta^3(\boldsymbol{p}_3-\boldsymbol{p}_3^{\prime}-\boldsymbol{p}_4-\boldsymbol{p}_5)  \Bigg\{\Big[\langle s_3^{\prime}|I|s_3\rangle\langle s_5\bar{s}_4|\boldsymbol{\sigma}|0\rangle-\langle s_3^{\prime}|\boldsymbol{\sigma}|s_3\rangle\langle s_5\bar{s}_4|I|0\rangle\Big] \nonumber \\
&& \!\!\!\!\! \cdot \Big[\Big(\frac{\boldsymbol{p}_5}{2m_5}+\frac{\boldsymbol{p}_4}{2m_4}\Big)-\Big(\frac{\boldsymbol{p}_3^{\prime}}{2m_3^{\prime}}+\frac{\boldsymbol{p}_3}{2m_3}\Big)\Big]  +i\langle s_3^{\prime}|\boldsymbol{\sigma}|s_3\rangle\times\langle s_5\bar{s}_4|\boldsymbol{\sigma}|0\rangle\cdot\Big[\Big(\frac{\boldsymbol{p}_4}{2m_4}-\frac{\boldsymbol{p}_5}{2m_5}\Big)-\Big(\frac{\boldsymbol{p}_3}{2m_3}-\frac{\boldsymbol{p}_3^{\prime}}{2m_3^{\prime}}\Big)\Big]\Bigg\}, \label{eq:1to3PCH_W} \\
\hat{H}_{W,1\to 3}^{(\text{PV})} &=& \frac{G_F}{\sqrt{2}}\frac{\beta V_{ud}V_{us}}{(2\pi)^3}\hat{\alpha}_3^{(+)}\hat{I}_{P}\frac{\delta_{c_4c_5}\delta_{c_3c_3^{\prime}}}{\sqrt{3}}\delta^3(\boldsymbol{p}_3-\boldsymbol{p}_3^{\prime}-\boldsymbol{p}_4-\boldsymbol{p}_5) \Big(-\langle s_3^{\prime}|I|s_3\rangle\langle s_5\bar{s}_4|I|0\rangle+\langle s_3^{\prime}|\boldsymbol{\sigma}|s_3\rangle\cdot\langle s_5\bar{s}_4|\boldsymbol{\sigma}|0\rangle\Big),     \label{eq:1to3PVH_W}
    \end{eqnarray}
\end{widetext}
where $m_i$, $s_i$ and $c_i$ are the constituent mass, spin and color of the $i$-th quark, respectively. And $\bar{s}_4$ stands for the spin of 4-th antiquark. The Pauli operator $\boldsymbol{\sigma}$ and the momentum $\boldsymbol{p}_i$ are the spin operator and spatial operator acting on the spin wave function and spatial wave function, respectively. In order to evaluate the spin matrix element including an antiquark, the particle-hole conjugation~\cite{Racah:1942gsc} is employed. With the particle-hole conjugation relation $|j,-m\rangle\to(-1)^{j+m}|j,m\rangle$, the antiquark spin transforms as follows: $\langle\bar{\uparrow}|\to|\downarrow\rangle$ and $\langle\bar{\downarrow}|\to-|\uparrow\rangle$. $\delta_{c_4c_5}\delta_{c_3c_3^{\prime}}/\sqrt{3}$ is the color operator for the transition vertex $q_3\to q_3^{\prime}\bar{q}_4q_5$, reflecting that the $W$ boson does not carry color, and the factor $1/\sqrt{3}$ originates from the normalization coefficient of the color singlet $q_5\bar{q}_4$ pair produced from the $W$ boson. It is interesting to note that the factor $1/\sqrt{3}$ has been regarded as a requirement for the quark pair creation (QPC) model which characterizes the vacuum to the normalized color-anticolor state~\cite{Micu:1968mk, LeYaouanc:1972vsx, LeYaouanc:1973ldf, LeYaouanc:1974cvx, LeYaouanc:1977gm}, and the QPC model has been successfully applied to many strong decays~\cite{Blundell:1996as, Lu:2016bbk, Chen:2007xf, Chen:2016iyi, Zhao:2016qmh, Gui:2018rvv, Deng:2023mza}. Extensively, the factor $1/\sqrt{3}$ can also be found in quark model calculations about the vacuum-to-meson form factors~\cite{Chernyak:1983ej, Chernyak:1981zz, Cheng:2009yz}.

In Eqs.~\eqref{eq:1to3PCH_W} and \eqref{eq:1to3PVH_W}, $\hat{\alpha}^{(+)}$ is the flavor-changing operator that converts an $s$ quark to a $u$ quark, and $\hat{I}_{P}$ is the isospin operator for the pseudoscalar meson ($\pi$ and $K$) production process. If we choose the de Swart’s flavor convention~\cite{deSwart:1963pdg, Lu:2024ajt}, $\hat{I}_{P}^{j}$ has the form of 
\begin{equation}
    \hat{I}^j_{P}=\left\{
    \begin{array}{ll}
        b_{d,j}^{\dagger}b_{s,j}, &\text{for}\ K^-,\\
        -\frac{1}{\sqrt{2}}b_{d,j}^{\dagger}b_{u,j}, &\text{for}\ \pi^0,
    \end{array}
    \right.
\end{equation}
for the singly Cabibbo-suppressed processes and will act on the $j$-th quark of the initial baryon after considering the flavor wave function of the emitted $\pi$ and $K$. Combining the effects of flavor changing and $\pi$ and $K$ production, the flavor operator can be expressed as
\begin{equation}
    \hat{\mathcal{O}}_{\text{flavor}}=\left\{
    \begin{array}{ll}
        -b_{u,3}^{\dagger}b_{s,3}, &\Omega^-\to \Xi^0\pi^-\  [\text{DPE}],\\
        (b_{u,3}^{\dagger}b_{s,3})(-\frac{1}{\sqrt{2}}b_{d,3}^{\dagger}b_{u,3}), &\Omega^-\to \Xi^-\pi^0\ [\text{CS-1}],\\
        (b_{u,3}^{\dagger}b_{s,3})(b_{d,1}^{\dagger}b_{s,1}), &\Omega^-\to \Lambda K^-\ [\text{CS-2}].
        \nonumber
    \end{array}
    \right.
\end{equation}

Note that Eqs.~\eqref{eq:1to3PCH_W} and \eqref{eq:1to3PVH_W} present the effective operator for only one specific interaction vertex structure. In the complete calculation, it is necessary to sum over all different identical contributions to ensure that all allowed interaction processes are fully taken into account. The symmetry factor $\beta$ is introduced to account for the multiple contributions from the interaction vertices. Since the $\text{SU}(6)\otimes\text{O}(3)$ wave functions of baryons are fully symmetric in flavor, spin, and space under interchange of any two quarks, thus we have
\begin{equation}
\langle\mathbb{B}_f|\sum_{j}\hat{H}_{W,s_j\to u_jd_5\bar{u}_4}^{(\text{P})}|\mathbb{B}_i\rangle =   \beta\langle\mathbb{B}_f|\hat{H}_{W,s_3\to u_3\bar{u}_4d_5}^{(\text{P})}|\mathbb{B}_i\rangle, \nonumber
\end{equation}
which will simplify the calculation. If the color factor is absorbed into the definition of the symmetry factor, then $\beta$ equals $A_3^1 \times 1 = 3$ and $A_3^1 \times 1/3 = 1$ for the DPE and CS-1 processes, respectively. This is because, in these two processes, there are two $s$ quarks in $\Omega$ are the spectator quarks, and the subsequent factors of 1 and 1/3 correspond to their respective color factors. However, $\beta=A_3^2\times 1/3=2$ in the CS-2 processes since only one $s$ quark in $\Omega$ is the spectator quark and the subsequent suppression factor is 1/3.

Then, the transition amplitudes of the DPE and CS processes shown in Fig.~\ref{fig:quark-level diagram} can be expressed as
\begin{widetext}
\begin{eqnarray}
\mathcal{M}_{\text{DPE/CS}}^{J_f,J_f^z;J_i,J_i^z} &=& \langle \mathbb{B}_{f}(\boldsymbol{P}_f,J_f,J_f^z)\mathbb{M}(\boldsymbol{k})|\hat{H}_{W,1\to3}^{\text{(P)}}|\mathbb{B}_i(\boldsymbol{P}_i,J_i,J_i^z)\rangle \nonumber \\
        &=& \sum_{S_f^z,L_f^z;S_i^z,L_i^z}\langle \phi_f|\hat{\mathcal{O}}_{\text{flavor}}|\phi_i\rangle\langle \chi_{S_f,S_f^z}|\hat{\mathcal{O}}_{\text{spin}}|\chi_{S_i,S_i^z}\rangle  \langle\psi_{P}(\boldsymbol{k})\psi_{N_fL_fL_f^z}(\boldsymbol{P}_f)|\hat{\mathcal{O}}_{\text{spatial}}(\boldsymbol{p_i})|\psi_{N_iL_iL_i^z}(\boldsymbol{P}_i)\rangle \nonumber \\
        & =& \sum_{S_f^z,L_f^z;S_i^z,L_i^z}\langle \phi_f|\hat{\mathcal{O}}_{\text{flavor}}|\phi_i\rangle\langle \chi_{S_f,S_f^z}|\hat{\mathcal{O}}_{\text{spin}}|\chi_{S_i,S_i^z}\rangle  I_{\text{\text{DPE/CS}}}^{L_f,L_f^z;L_i,L_i^z}.
        \label{eq:DPE/CS amp}
 \end{eqnarray}
 \end{widetext}
 
Apart from the color factor, the quark arrangements in the DPE and CS make a difference between these two processes via the spatial integrals. For the DPE process, the momentum conservation requires $\boldsymbol{P}_f=\boldsymbol{p}_1+\boldsymbol{p}_2+\boldsymbol{p}_3^{\prime}$ and $\boldsymbol{k}=\boldsymbol{p}_5+\boldsymbol{p}_4$, while $\boldsymbol{P}_f=\boldsymbol{p}_1+\boldsymbol{p}_2+\boldsymbol{p}_5$ and $\boldsymbol{k}=\boldsymbol{p}_3^{\prime}+\boldsymbol{p}_4$ for the CS-1 process; $\boldsymbol{P}_f=\boldsymbol{p}_5+\boldsymbol{p}_2+\boldsymbol{p}_3^{\prime}$ and $\boldsymbol{k}=\boldsymbol{p}_1+\boldsymbol{p}_4$ for the CS-2 process. They are guaranteed by the delta function in the following expressions of the spatial wave function convolution $I_{\text{DPE/CS}}^{L_f,L_f^z;L_i,L_i^z}$.

\begin{table}[htbp]
    \caption{The flavor matrix elements in the DPE and CS processes for $\Omega^-$ decays.}
    \centering
    \label{tab:DPE and CS fla}
    \begin{tabular}{ccc}
    \hline\hline
    Decay channels  &$\langle \phi_f^{\lambda}|\hat{O}_{\text{flavor}}|\phi_i^s\rangle$ &$\langle \phi_f^{\rho}|\hat{O}_{\text{flavor}}|\phi_i^s\rangle$\\
    \hline
    $\Omega^-\to\Xi^0\pi^-$ (DPE)  &$-\frac{2}{\sqrt{6}}$  &0  \\
    $\Omega^-\to\Xi^-\pi^0$ (CS-1)  &$-\frac{1}{\sqrt{3}}$  &0  \\
    $\Omega^-\to\Lambda K^-$ (CS-2)  &$-\frac12$ &$-\frac{1}{2\sqrt{3}}$\\
    \hline\hline
    \end{tabular}
\end{table}

The flavor and spin matrix elements in Eq.~(\ref{eq:DPE/CS amp}) have different results for the DPE and CS processes. They are listed separately in Tabs.~\ref{tab:DPE and CS fla} and \ref{tab:DPE and CS spin}. It is noteworthy that the evaluated matrix elements of the $\hat{\mathcal{O}}_{\text{spin}}^{(\text{PV})}$ operator vanish in the present calculation. This is because $\hat{H}_{W,1\to 3}^{(\text{PV})}$ (see Eq.~(\ref{eq:1to3PVH_W})) we used has no momentum dependence and thus acts an $S$-wave operator. For the weak decay $3/2^+\to 1/2^++0^+$, however, angular momentum and parity conservation allow a $P$-wave (PC) or $D$-wave (PV) configuration between the final-state particles.

\begin{table}[htbp]
    \centering
    \caption{The spin matrix elements for the PC transitions in the DPE and CS processes.}
    \scalebox{0.88}{
    \begin{tabular}{c|ccc|ccc}
        \hline\hline
        \multirow{2}{*}{$\hat{\mathcal{O}}_{\text{spin}}^{(\text{PC})}$}&\multicolumn{3}{c|}{$\langle \chi_{\frac12,-\frac12}^{\lambda}|\hat{\mathcal{O}}_{\text{spin}}^{(\text{PC})}|\chi_{\frac32,-\frac12}^{s}\rangle$ }&\multicolumn{3}{c}{$\langle \chi_{\frac12,-\frac12}^{\rho}|\hat{\mathcal{O}}_{\text{spin}}^{(\text{PC})}|\chi_{\frac32,-\frac12}^{s}\rangle$ }\\\cline{2-7}
         &DPE &CS-1 &CS-2 &DPE &CS-1 &CS-2 \\
         \hline
         $\langle s_3^{\prime}|I|s_3\rangle\langle s_5\bar{s}_4|\bs{\sigma}|0\rangle$ &0&$\frac23\hat{\boldsymbol{k}}$&$-\frac13\hat{\boldsymbol{k}}$&0&0&$-\frac{1}{\sqrt{3}}\hat{\boldsymbol{k}}$\\
        $\langle s_3^{\prime}|\bs{\sigma}|s_3\rangle\langle s_5\bar{s}_4|I|0\rangle$  &$\frac43\hat{\boldsymbol{k}}$&$\frac23\hat{\boldsymbol{k}}$&$\frac23\hat{\boldsymbol{k}}$&0&0&0\\
        $\langle s_3^{\prime}|\boldsymbol{\sigma}|s_3\rangle\times\langle s_5\bar{s}_4|\boldsymbol{\sigma}|0\rangle$&0&$-\frac43i\hat{\boldsymbol{k}}$&$i\hat{\boldsymbol{k}}$&0&0&$\frac{1}{\sqrt{3}}i\hat{\boldsymbol{k}}$\\
        \hline\hline
    \end{tabular}}
    \label{tab:DPE and CS spin}
\end{table}

The spatial matrix elements can be calculated by the overlap integral of the initial-state wave function acting under the spatial operator with the final-state wave function. Taking the CS-1 process as an example, the expression is given by:
\begin{widetext}
\begin{eqnarray}
        I_{\text{CS-1}}^{L_f,L_f^z;L_i,L_i^z} &=& \langle \psi_{P}(\boldsymbol{k})\psi_{N_fL_fL_f^z}(\boldsymbol{P}_f)|\hat{\mathcal{O}}_{W,1\to 3}^{\text{spatial}}(\boldsymbol{p}_i)|\psi_{N_iL_iL_i^z}(\boldsymbol{ P}_i)\rangle  = \int d\boldsymbol{p}_1d\boldsymbol{p}_2d\boldsymbol{p}_3d\boldsymbol{p}_1^{\prime}d\boldsymbol{p}_2^{\prime}d\boldsymbol{p}_3^{\prime}d\boldsymbol{p}_4d\boldsymbol{p}_5 \delta^3(\boldsymbol{p}_1^{\prime}-\boldsymbol{p}_1) \times \nonumber \\
        && \delta^3(\boldsymbol{p}_2^{\prime}-\boldsymbol{p}_2)\delta^3(\boldsymbol{p}_3-\boldsymbol{p}_3^{\prime}-\boldsymbol{p}_4-\boldsymbol{p}_5) \psi_{N_fL_fL_f^z}^*(\boldsymbol{p}_1^{\prime},\boldsymbol{p}_2^{\prime},\boldsymbol{p}_5) \psi_P^*(\boldsymbol{p}_3^{\prime},\boldsymbol{p}_4)  \hat{\mathcal{O}}_{W,1\to 3}^{\text{spatial}}(\boldsymbol{p}_i)\psi_{N_iL_iL_i^z}(\boldsymbol{p}_1,\boldsymbol{p}_2,\boldsymbol{p}_3) \times \nonumber \\
        &&\delta^3(\boldsymbol{P}_i-\boldsymbol{p}_1-\boldsymbol{p}_2-\boldsymbol{p}_3)\delta^3(\boldsymbol{k}-\boldsymbol{p}_3^{\prime}-\boldsymbol{p}_4)\delta^3(\boldsymbol{P}_f-\boldsymbol{p}_1^{\prime}-\boldsymbol{p}_2^{\prime}-\boldsymbol{p}_5),
\end{eqnarray}
\end{widetext}
where $\hat{\mathcal{O}}_{W,1\to3}^{\text{spatial}}(\boldsymbol{p}_i)$ is a function of the quark momentum $\boldsymbol{p}_i$, such as $\boldsymbol{p}_5/(2m_5)+\boldsymbol{p}_4/(2m_4)$ or just 1 for $\hat{H}_{W,1\to 3}^{(\text{PV})}$. The significant difference between the two integral functions \(I_{\text{DPE}}^{L_f,L_f^z;L_i,L_i^z}\) and \(I_{\text{CS}}^{L_f,L_f^z;L_i,L_i^z}\) is the different momentum conservation conditions. This indicates that, for the DPE and CS processes, apart from the different color factor, there are also differences in the convolution of their spatial wave functions. This is a dynamic feature with the quark model approach. The detailed expressions for the amplitudes of the DPE and CS processes in each hyperon decay channel can be found in Appendix \ref{app:amp}.

\subsection{The decay amplitudes of $su\to ud$ processes}\label{sec-1 to 3}

The hyperon two-body hadronic weak decays also involve a typical transition processes known as the pole terms which describe the $su\to ud$ internal conversion mechanism via $W$-boson exchange. As shown in Fig.~\ref{fig:quark-level diagram}, only the $\Lambda K^-$ channel contains such a pole process where the strong interaction that emits a kaon meson first occurs, followed by the weak interaction that changes the flavor. For the pole terms, the general expression of decay amplitude can be written as~\cite{LeYaouanc:1978ef, Richard:2016hac, Niu:2020gjw}
\begin{widetext}
\begin{eqnarray}
\mathcal{M}_{\text{PT(PC)}}^{J_f,J_f^z;J_i,J_i^z} &=& \sum_{\mathbb{B}_m}\langle\mathbb{B}_f(\boldsymbol{P}_f;J_f,J_f^z)|\frac{i\hat{H}_{W,2\to2}^{\text{(PC)}}}{\slashed{p}_{\mathbb{B}_m}-m_{\mathbb{B}_m}+i\frac{\Gamma_{\mathbb{B}_m}}{2}}|\mathbb{B}_m(\boldsymbol{P}_f;J_f,J_f^z)\rangle\langle\mathbb{B}_m(\boldsymbol{P}_f;J_f,J_f^z)|\hat{H}_{P}|\mathbb{B}_i(\boldsymbol{P}_i;J_i,J_i^z)\rangle,\\
\mathcal{M}_{\text{PT(PV)}}^{J_f,J_f^z;J_i,J_i^z} & =& \sum_{\mathbb{B}_m^{\prime}}\langle\mathbb{B}_f(\boldsymbol{P}_f;J_f,J_f^z)|\frac{i\hat{H}_{W,2\to2}^{\text{(PV)}}}{\slashed{p}_{\mathbb{B}_m^{\prime}}-m_{\mathbb{B}_m^{\prime}}+i\frac{\Gamma_{\mathbb{B}_m^{\prime}}}{2}}|\mathbb{B}_m^{\prime}(\boldsymbol{P}_f;J_f,J_f^z)\rangle\langle\mathbb{B}_m^{\prime}(\boldsymbol{P}_f;J_f,J_f^z)|\hat{H}_{P}|\mathbb{B}_i(\boldsymbol{P}_i;J_i,J_i^z)\rangle,    \label{eq:PT amp}
\end{eqnarray}
\end{widetext}
where $\mathbb{B}_m$ and $\mathbb{B}_m^{\prime}$ denote the intermediate baryon states with quantum numbers of $J^P=1/2^+$ and $1/2^-$, respectively. The dominant contributions of the pole terms come from the processes that the intermediate states are approximately on-shell. Thus, the approximation of propagator $1/(\slashed{p}-m+i\Gamma/2) \approx 2m/(p^2-m^2+i m\Gamma)$ is applied. This treatment will bring uncertainties into the theoretical results since the intermediate states are generally off-shell. Nevertheless, such uncertainties can be absorbed into the quark model parameters for which the range of the favored values by experimental data can be estimated.

In this work, the intermediate state is identified as the $\Xi$ resonance with the flavor component $uss$. There are only two $\Xi$ baryons with 4-star ratings whose spin and parity are known, as reported in the PDG~\cite{ParticleDataGroup:2024cfk}. They are the octet ground states $\Xi(1315)^0$ and $\Xi(1322)^-$ with $J^P=1/2^+$ ($1^2S_{1/2^+}$), and the decuplet ground state $\Xi(1535)$ with $J^P=3/2^+$ ($1^4S_{3/2^+}$). However, the low-lying excited $\Xi$ states with negative parity have not been determined experimentally. There are two candidates listed in the PDG, one is the 2-star state $\Xi(1620)$ with mass about 1620 MeV and width of $32_{-9}^{+8}$ MeV, and the other one is 3-star state $\Xi(1690)$ with mass of $1690 \pm 10$ MeV and width of $20 \pm 15$ MeV. The BaBar collaboration, in their analysis of the decay $\Lambda_c^+\to\Xi^-\pi^+K^+$ and quantitative description of the $\Xi(1530)^0$ line shape, determined the $J^P$ of $\Xi(1690)$ to be $1/2^-$~\cite{BaBar:2008myc}. Furthermore, the BESIII collaboration observed the $\Xi(1690)^-$ state in $K^-\Lambda$ spectrum and confirmed its $J^P$ as $1/2^-$ through a partial wave analysis~\cite{BESIII:2023mlv}. This $J^P$ assignment is consistent with that of BaBar, but the reported width of $81_{-9}^{+10}$ MeV is larger. The evidence for the lower-mass $\Xi(1620)$ has been rather weak until recently. On the theoretical side, quark-model studies offer predictions for $\Xi(1/2^-)$ states. Ref.~\cite{Xiao:2013xi} assigned the $\Xi(1690)$ to the first orbital excitation $1^2P_{1/2^-}$, based on calculations of its partial decay widths in the chiral quark model (CQM). Similarly, Pervin and Roberts proposed that $\Xi(1690)$ could be identified as the first orbital excitation with $J^P=1/2^-$, although their model predicted a mass about 35 MeV heavier than the $\Xi(1690)$~\cite{Pervin:2007wa}. They believed that a more microscopic treatment of spin-orbit interactions, can be expected to drive this state to slightly lower mass. Consequently, for the pole term processes, we adopt the following treatment: the nonfactorizable PV amplitude primarily arises from the first orbital excitation state $1^2P_{1/2^-}$, while the nonfactorizable PC amplitude is determined by the ground state $1^2S_{1/2^+}$.

The CQM has been often applied to the production of light pseudoscalar mesons in various processes~\cite{Zhong:2008kd, Ni:2023lvx}. In this work, for the strong interaction vertex of the kaon emission that occurred first as shown in Fig.~\ref{fig:quark-level diagram} (d), we employ the CQM to study it with the transition Hamiltonian given by:
\begin{align}
    \begin{split}
        \hat{H}_{P}=\frac{1}{f_{P}}\sum_j\bar{\psi}_j\gamma_{\mu}\gamma_5\partial^{\mu}\phi_{P}\hat{I}_j^{P}\psi_j,
    \end{split}
\end{align}
where $f_{P}$ is the decay constant of the pseudoscalar meson, and $j$ denotes the $j$-th quark inside the baryon that interacts with the emitted pseudoscalar meson. In momentum representation, the chiral Lagrangian for quark-pseudoscalar meson pseudovector coupling reduces to the non-relativistic form up to order of $1/m$ as:   
\begin{eqnarray}
        \hat{H}_P^{NR} &=& \frac{1}{f_P\sqrt{(2\pi)^32\omega}}\sum_j\Big[\frac{\omega}{2m_j}\boldsymbol{\sigma}_j\cdot\boldsymbol{p}_j+\frac{\omega}{2m_j^{\prime}}\boldsymbol{\sigma}_j\cdot\boldsymbol{p}_j^{\prime} \nonumber \\
        & & -\boldsymbol{\sigma}_j\cdot\boldsymbol{k}\Big]\hat{I}_j^P\delta^3(\boldsymbol{p}_j-\boldsymbol{p}_j^{\prime}-\boldsymbol{k}),
        \label{eq:CQM L}
 \end{eqnarray}
where $\omega$ and $\boldsymbol{k}$ are the energy and three-momentum of the final-state pseudoscalar meson in the initial-state rest frame; $\boldsymbol{p}_j$ and $\boldsymbol{p}_j^{\prime}$ are the three-momenta of the $j$-th quark before and after emitting the pseudoscalar meson. For the pole term contribution to the $\Omega^- \to \Lambda K^-$ decay, we take the kaon decay constant $f_K=159$ MeV and $\hat{I}_j^{K^-}=a_j^{\dagger}(u)a_j(s)$ is the isospin operator for the $K^-$ production with $a_j^{\dagger}(u)$ and $a_j(s)$ representing the creation and annihilation operators for quarks. This isospin operator acts on the flavor wave function, and the only non-vanishing flavor matrix elements is $\langle\phi_{\Xi^0}^{\lambda}|\hat{I}_3^{K^-}|\phi_{\Omega^-}^s\rangle=2/\sqrt{6}$.

\begin{table}[htbp]
    \centering
    \caption{Values of the reduced matrix elements for each type of vector operator $\tau$ and the subscript number $i$ ($i=1,2,3$) indicates that the operator acts on the $i$-th quark.}
    \label{tab:spin}
    \scalebox{1}{
    \begin{tabular}{c|ccccc}
    \hline\hline
    $\tau$& $\sigma_1$ & $\sigma_2$ & $\sigma_3$ & $\sigma_2\times\sigma_3$ & $\sigma_1\times\sigma_2$\\
    \hline
    $\langle\chi_{\frac12}^{\rho}||\tau^{(1)}||\chi_{\frac12}^{\rho}\rangle$ 
    & $0$ & $0$ & $\sqrt{6}$ & $0$ &$0$\\
    $\langle\chi_{\frac12}^{\lambda}||\tau^{(1)}||\chi_{\frac12}^{\lambda}\rangle$ 
    & $\frac{2\sqrt{6}}{3}$ & $\frac{2\sqrt{6}}{3}$ & $-\frac{\sqrt{6}}{3}$ & $0$ &$0$\\
    $\langle\chi_{\frac12}^{\rho}||\tau^{(1)}||\chi_{\frac12}^{\lambda}\rangle$ 
    & $-\sqrt{2}$ & $\sqrt{2}$ & $0$ & $-2\sqrt{2}i$ &$-2\sqrt{2}i$\\
    $\langle\chi_{\frac32}^{s}||\tau^{(1)}||\chi_{\frac12}^{\rho}\rangle$ 
    & $2$ & $-2$ & $0$ & $2i$ &$-4i$\\
    $\langle\chi_{\frac32}^{s}||\tau^{(1)}||\chi_{\frac12}^{\lambda}\rangle$ 
    & $\frac{2\sqrt{3}}{3}$ & $\frac{2\sqrt{3}}{3}$ & $-\frac{4\sqrt{3}}{3}$ & $-2\sqrt{3}i$ &$0$\\
    $\langle\chi_{\frac32}^{s}||\tau^{(1)}||\chi_{\frac32}^{s}\rangle$ 
    & $\frac{2\sqrt{15}}{3}$ & $\frac{2\sqrt{15}}{3}$ & $\frac{2\sqrt{15}}{3}$ & $0$ &$0$\\
    \hline\hline
    \end{tabular}}
\end{table}

Since the baryon wave function is fully symmetric with the asymmetric color wave function, the strong interaction matrix element can be simplified using $\langle\mathbb{B}_f|\sum_{j=1}^3\hat{H}_{Pj}^{NR}|\mathbb{B}_i\rangle=3\langle\mathbb{B}_f|\hat{H}_{P(j=3)}^{NR}|\mathbb{B}_i\rangle$. The Pauli operator and momentum operator in Eq.~(\ref{eq:CQM L}) are coupled together, and we handle them using the following formula:
\begin{eqnarray}
\langle\boldsymbol{\sigma}\cdot\boldsymbol{p}\rangle &=& \langle\chi_{s^{\prime}s_z^{\prime}}^{\nu^{\prime}}|\boldsymbol{\sigma}|\chi_{ss_z}^{\nu}\rangle\cdot\boldsymbol{p} = \sqrt{\frac{4\pi}{3}}|\boldsymbol{p}| \times \nonumber \\
        && \sum_{m=-1}^1(-1)^m\langle\chi_{s^{\prime}s_z^{\prime}}^{\nu^{\prime}}|\sigma_{m}^{(1)}|\chi_{ss_z}^{\nu}\rangle Y_{1-m}(\hat{p}),
        \label{eq:sigma cdot p}
\end{eqnarray}
where $Y_{1m}$ is the spherical harmonic function; $\chi_{s s_z}^{\nu}$ represents the spin wave function, where the superscript $\nu = s, \rho, \lambda$ denotes its permutation symmetry and the subscript corresponds to the spin quantum number and its projection; $\sigma_m^{(1)}$ is the component of the first-rank irreducible tensor operator $\boldsymbol{\sigma}$:
\begin{align}
    \begin{split}
    \sigma_0^{(1)}=\sigma_z,\ \ \ \sigma_{\pm 1}^{(1)}=\mp\frac{1}{\sqrt{2}}(\sigma_x\pm i\sigma_y).
    \end{split}
\end{align}
According to the Wigner-Eckart theorem, the matrix elements of $\sigma_m^{(1)}$ acting on the spin wave functions can be factorized as
\begin{eqnarray}
 \langle\chi_{s^{\prime}s_z^{\prime}}^{\nu^{\prime}}|{\sigma}_m^{(1)}|\chi_{ss_z}^{\nu}\rangle = \langle s,s_z;1,m|s^{\prime},s_z^{\prime}\rangle\frac{\chi^{\nu^{\prime}\nu}_{s^{\prime}s}}{\sqrt{2s^{\prime}+1}},
        \label{eq:Winger}
   \end{eqnarray}
with $\chi^{\nu^{\prime}\nu}_{s^{\prime}s} = \langle\chi_{s^{\prime}}^{\nu^{\prime}}||\sigma^{(1)}||\chi_s^{\nu}\rangle$ the reduced matrix element. With all independent reduced matrix elements known, spin matrix elements for all arbitrary spin combinations can be computed via the Hermiticity relation: $\chi^{\nu^{\prime}\nu}_{s^{\prime}s} = (-1)^{s-s^{\prime}} \langle\chi_{s}^{\nu}||\sigma^{(1)}||\chi_{s^{\prime}}^{\nu^{\prime}}\rangle$. An extra global minus sign is needed for anti-Hermitian operator like $\sigma \times \sigma$. The six independent reduced matrix elements and the values for each type of vector operator are listed in Tab.~\ref{tab:spin}.

Finally, for the strong matrix element in Eq. (\ref{eq:PT amp}), we have
\begin{widetext}
\begin{eqnarray}
&&  \langle\mathbb{B}_m(\boldsymbol{P}_f;J_f,J_f^z)|\hat{H}_P|\mathbb{B}_i(\boldsymbol{P}_i;\boldsymbol{J}_i,J_i^z)\rangle = \frac{3}{f_P\sqrt{(2\pi)^32\omega}}\sum_{S_f^z,L_f^z;S_i^z,L_i^z}\langle \phi_f|\hat{I}_3^P|\phi_i\rangle\langle \chi_{S_f,S_f^z}|\hat{\mathcal{O}}_{\text{spin}}(\sigma_{3z},\sigma_3^+,\sigma_3^-)|\chi_{S_i,S_i^z}\rangle \nonumber \\
        && \times\int d\boldsymbol{p}_1d\boldsymbol{p}_2d\boldsymbol{p}_3d\boldsymbol{p}_1^{\prime}d\boldsymbol{p}_2^{\prime}d\boldsymbol{p}_3^{\prime}\delta^3(\boldsymbol{p}_1^{\prime}-\boldsymbol{p}_1)\delta^3(\boldsymbol{p}_2^{\prime}-\boldsymbol{p}_2)\delta^3(\boldsymbol{p}_3^{\prime}-\boldsymbol{p}_3+\boldsymbol{k}) \psi_{N_fL_fL_f^z}^*(\boldsymbol{p}_1^{\prime},\boldsymbol{p}_2^{\prime},\boldsymbol{p}_3^{\prime})\hat{\mathcal{O}}_{P}^{\text{spatial}}(\boldsymbol{p}_{3z},\boldsymbol{p}_{3}^+,\boldsymbol{p}_{3}^-) \nonumber \\
        && \times \psi_{N_iL_iL_i^z}(\boldsymbol{p}_1,\boldsymbol{p}_2,\boldsymbol{p}_3)\delta^3({\boldsymbol{P}}_i-\boldsymbol{p}_1-\boldsymbol{p}_2-\boldsymbol{p}_3)\delta^3({\boldsymbol{P}}_f-\boldsymbol{p}_1^{\prime}-\boldsymbol{p}_2^{\prime}-\boldsymbol{p}_3^{\prime}).
\end{eqnarray}
\end{widetext}

Subsequently, for the $su\to ud$ internal conversion process as shown in Fig.~\ref{fig:quark-level diagram} (d), the non-relativistic Hamiltonian can be written as~\cite{Richard:2016hac, Niu:2020gjw}:
\begin{widetext}
    \begin{align}
    \begin{split}
        \hat{H}_{W,2\to 2}^{(\text{PC})}&=\frac{G_F}{\sqrt{2}}\sum_{i\neq j}\frac{V_{q_iq_i^{\prime}}V_{q_jq_j^{\prime}}}{(2\pi)^3}{\delta_{c_ic_i^{\prime}}\delta_{c_jc_j^{\prime}}}\hat{\alpha}_i^{(-)}\hat{\beta}_j^{(+)}\delta^3(\boldsymbol{p}_i^{\prime}+\boldsymbol{p}_j^{\prime}-\boldsymbol{p}_i-\boldsymbol{p}_j)\Big(\langle{s_i}^{\prime}|I|{s_i}\rangle\langle{s_j^{\prime}}|I|{s_j}\rangle-\langle{s_i}^{\prime}|\boldsymbol{\sigma}_i|{s_i}\rangle\cdot\langle{s_j^{\prime}}|\boldsymbol{\sigma}_j|{s_j}\rangle\Big),\\
        \hat{H}_{W,2\to 2}^{(\text{PV})}&=\frac{G_F}{\sqrt{2}}\sum_{i\neq j}\frac{V_{q_iq_i^{\prime}}V_{q_jq_j^{\prime}}}{(2\pi)^3}{\delta_{c_ic_i^{\prime}}\delta_{c_jc_j^{\prime}}}\hat{\alpha}_i^{(-)}\hat{\beta}_j^{(+)}\delta^3(\boldsymbol{p}_i^{\prime}+\boldsymbol{p}_j^{\prime}-\boldsymbol{p}_i-\boldsymbol{p}_j)\\
        &\times \Bigg\{\Big(\langle{s_i}^{\prime}|I|{s_i}\rangle\langle{s_j^{\prime}}|\boldsymbol{\sigma}_j|{s_j}\rangle-\langle{s_i}^{\prime}|\boldsymbol{\sigma}_i|{s_i}\rangle\langle{s_j^{\prime}}|I|{s_j}\rangle\Big)\cdot\Big[\Big(\frac{\boldsymbol{p}_i}{2m_i}-\frac{\boldsymbol{p}_j}{2m_j}\Big)+\Big(\frac{\boldsymbol{p}_i^{\prime}}{2m_i^{\prime}}-\frac{\boldsymbol{p}_j^{\prime}}{2m_j^{\prime}}\Big)\Big]\\
        &+i\Big(\langle{s_i}^{\prime}|\boldsymbol{\sigma}_i|{s_i}\rangle\times\langle{s_j^{\prime}}|\boldsymbol{\sigma}_j|{s_j}\rangle\Big)\cdot\Big[\Big(\frac{\boldsymbol{p}_i}{2m_i}-\frac{\boldsymbol{p}_j}{2m_j}\Big)-\Big(\frac{\boldsymbol{p}_i^{\prime}}{2m_i^{\prime}}-\frac{\boldsymbol{p}_j^{\prime}}{2m_j^{\prime}}\Big)\Big]
        \Bigg\},
        \label{eq:2to2 H_W}
    \end{split}
\end{align}
\end{widetext}
where the subscripts $i$ and $j$ ($i,j=1,2,3$ and $i\neq j$) indicate the quarks experiencing the weak interaction; $\hat{\alpha}_i^{(-)}$ and $\hat{\beta}_j$ are the flavor-changing operators, namely, $\alpha_i^{(-)}u_j=\delta_{ij}d_i$, $\hat{\beta}_j^{(+)}s_i=\delta_{ij}u_j$. The {$\delta_{c_ic_i^{\prime}}\delta_{c_jc_j^{\prime}}$ is the color operator for the transition vertex $q_iq_j\to q_i^{\prime}q_j^{\prime}$, since the $W$ boson does not carry color.} In the calculation, we can fix the subscripts $i$ and $j$ to be 1 and 2 due to the symmetry of the total wave function, then the weak transition matrix element can be simplified to $\langle\mathbb{B}_f|\sum_{i\neq j}\hat{H}_{W,u_is_j\to d_iu_j}^{(\text{P})}|\mathbb{B}_i\rangle=3\langle\mathbb{B}_f|\hat{H}_{W,u_1s_2\to d_1u_2}^{(\text{P})}|\mathbb{B}_i\rangle$. Under this quark label, the non-vanishing flavor matrix elements are $\langle\phi_{\Lambda}^{\rho}|\hat{\alpha}_1^{(-)}\hat{\beta}_2^{(+)}|\phi_{\Xi^0}^{\lambda}\rangle = 1/(3\sqrt{2})$ and $\langle\phi_{\Lambda}^{\rho}|\hat{\alpha}_1^{(-)}\hat{\beta}_2^{(+)}|\phi_{\Xi^0}^{\rho}\rangle = 1/\sqrt{6}$. The matrix elements related to spin are handled in the same way as in Eqs.~(\ref{eq:sigma cdot p}) and (\ref{eq:Winger}), and their values are listed in Tab.~\ref{tab:spin}.

Finally, the weak matrix element for PC transition can be obtained by producing the integration as
\begin{widetext}    
\begin{eqnarray}
&& \langle\mathbb{B}_f(\boldsymbol{P}_f;J_f,J_f^z)|\hat{H}_{W,2,\to 2}^{(\text{PC})}|\mathbb{B}_m(\boldsymbol{P}_f;J_f,J_f^z)\rangle = \frac{G_F}{\sqrt{2}}\frac{6V_{ud}V_{us}}{(2\pi)^3}\sum_{S_f^z,L_f^z;S_i^z,L_i^z}\langle \phi_f|\hat{\alpha}_1^{(-)}\hat{\beta}_2^{(+)}|\phi_i\rangle\langle \chi_{S_f,S_f^z}|1-\boldsymbol{\sigma}_1\cdot\boldsymbol{\sigma}_2|\chi_{S_i,S_i^z}\rangle \nonumber \\
        && \times\int d\boldsymbol{p}_1d\boldsymbol{p}_2d\boldsymbol{p}_3d\boldsymbol{p}_1^{\prime}d\boldsymbol{p}_2^{\prime}d\boldsymbol{p}_3^{\prime}\delta^3({\boldsymbol{p}_1^{\prime}+\boldsymbol{p}_2^{\prime}-\boldsymbol{p}_1-\boldsymbol{p}_2})\delta^3(\boldsymbol{p}_3^{\prime}-\boldsymbol{p}_3)\psi_{N_fL_fL_f^z}^{*}(\boldsymbol{p}_1^{\prime},\boldsymbol{p}_2^{\prime},\boldsymbol{p}_3^{\prime}) \psi_{N_iL_iL_i^z}(\boldsymbol{p}_1,\boldsymbol{p}_2,\boldsymbol{p}_3) \nonumber \\
        && \times \delta^3(\boldsymbol{P}_i-\boldsymbol{p}_1-\boldsymbol{p}_2-\boldsymbol{p}_3)\delta^3(\boldsymbol{P}_f-\boldsymbol{p}_1^{\prime}-\boldsymbol{p}_2^{\prime}-\boldsymbol{p}_3^{\prime}).
 \end{eqnarray}
\end{widetext}
Analogously, we also can obtain the weak transition matrix element $\langle\mathbb{B}_f(\boldsymbol{P}_f;J_f,J_f^z)|\hat{H}_{W,2,\to 2}^{(\text{PV})}|\mathbb{B}_m(\boldsymbol{P}_f;J_f,J_f^z)\rangle$.

In Refs.~\cite{Cao:2025kvs,Richard:2016hac}, cancellations emerge separately for Type-A and Type-B pole contributions in $\Lambda\to N\pi$ decays, despite the individually sizable magnitudes of the exclusive Type-A and Type-B amplitudes. By contrast, the $\Omega^- \to \Lambda K^-$ decay involves only one type of pole term, corresponding to a strong interaction first followed by weak decay. The large amplitude originating from this pole term naturally the experimental data that the $\Lambda K^-$ mode has the largest branching ratio. We note in advance that the pole terms is indeed crucial for the decay $\Omega^-\to \Lambda K^-$ in order to reproduce the branching ratio of the experiment. This observation not only solidifies the physical significance of pole terms but also highlights their essential role in unraveling the decay dynamics of $\Omega^-$ hyperon.

\subsection{The nonrelativistic potential and baryon wave functions} \label{sec-baryon wave functions}

In the previous theoretical works.~\cite{Cao:2023csx, Cao:2025kvs, Wang:2026ghd, Niu:2020gjw}, a simple harmonic oscillator (H.O.) wave function was introduced to describe the motion of constituent quarks inside hadrons. During this process, a phenomenological parameter $\alpha$ was required. To address this issue, we adopt a numerical wave function based on the support of hadron spectroscopy. Specifically, we determine the parameters of the potential model by fitting existing hadron spectra, and then obtain the complete hadron wave function by solving the Schr\"{o}dinger equation numerically. This approach yields a wave function that more faithfully represents the motion of constituent quarks and will reduce the uncertainty associated with the $\alpha$ parameter of the simple H.O. wave function. To obtain the wave function of baryons, we take a non-relativistic Hamiltonian
\begin{align}
    \begin{split}
        \hat{H}=\sum_{i=1}^3(\frac{\boldsymbol{p}_i^2}{2m_i}+m_i)+\sum_{i<j}V(r_{ij})+C_0,
    \end{split}
\end{align}
where $\boldsymbol{p}_i^2/({2m_i})$ is the kinetic energy of the $i$-th constituent quark with mass $m_i$ and momentum $\boldsymbol{p}_i$, $V_{ij}$ is the effective potential between the $i$-th and $j$-th quarks with a distance $r_{ij}\equiv |\boldsymbol{r}_i-\boldsymbol{r}_j|$. $C_0$ is the zero point energy. The effective potential can be decomposed into the spin-independent and spin-dependent parts,
\begin{align}
    \begin{split}
        V(r_{ij})=V^{corn}(r_{ij})+V^{sd}(r_{ij}).
    \end{split}
\end{align}
The spin-independent part $V^{corn}(r_{ij})$ is adopted the well-known Cornell form~\cite{Eichten:1978tg}, i.e.,
\begin{align}
    V^{corn}(r_{ij})=\frac{b_{ij}}{2}r_{ij}-\frac23\frac{\alpha_{ij}}{r_{ij}},
\end{align}
where the first term is the linear confinement potential $V^{conf}$, and the second term is the Coulomb-like potential $V^{coul}$ derived from the one-gluon-exchange (OGE) model~\cite{Capstick:1986ter, Godfrey:1985xj}. $b_{ij}$ and $\alpha_{ij}$ are the slope parameter of the confinement potential and strong coupling constant between the $i$-th and $j$-th quarks, respectively. While, the spin-dependent potentials in OGE model can be decomposed into
\begin{align}
    \begin{split}
        V^{sd}(r_{ij})=V^{SS}(r_{ij})+V^T(r_{ij})+V^{LS}(r_{ij}),
    \end{split}
\end{align}
where $V^{SS}$, $V^T$ and $V^{LS}$ stand for the spin-spin, tensor, and the spin-orbit potentials, respectively. The spin
-spin and tensor potentials are given by
\begin{align}
    \begin{split}
        &V^{SS}(r_{ij})=\frac{2\alpha_{ij}}{3}\left\{\frac{\pi}{2}\cdot\frac{\sigma_{ij}^3e^{-\sigma_{ij}^2r_{ij}^2}}{\pi^{3/2}}\cdot\frac{16}{3m_im_j}(\boldsymbol{S}_i\cdot\boldsymbol{S}_j)\right\},\\
        &{V}^T(r_{ij})=\frac{2\alpha_{ij}}{3m_im_jr_{ij}^3}\left\{\frac{3(\boldsymbol{S}_i\cdot\boldsymbol{r}_{ij})(\boldsymbol{S}_j\cdot\boldsymbol{r}_{ij})}{r_{ij}^2}-\boldsymbol{S}_i\cdot\boldsymbol{S}_j\right\}. \notag
    \end{split}
\end{align}
The spin-orbit potential contains a color-magnetic part and a Thomas-precession part~\cite{Capstick:1986ter, Zhong:2024mnt}
\begin{eqnarray}
        V^{LS}(r_{ij})=V_{ij}^{so(\nu)}+V_{ij}^{so(s)},
\end{eqnarray}
where the two parts $V_{ij}^{so(\nu)}$ and $V_{ij}^{so(s)}$ are
\begin{align}
    \begin{split}
        {V}_{ij}^{so(\nu)}&=\frac{1}{r_{ij}}\frac{dV_{ij}^{coul}}{dr_{ij}}\Bigg(\frac{\boldsymbol{r}_{ij}\times\boldsymbol{p}_i\cdot\boldsymbol{S}_i}{2m_i^2}-\frac{\boldsymbol{r}_{ij}\times\boldsymbol{p}_j\cdot\boldsymbol{S}_j}{2m_j^2}\\
        &-\frac{\boldsymbol{r}_{ij}\times\boldsymbol{p}_j\cdot\boldsymbol{S}_i-\boldsymbol{r}_{ij}\times\boldsymbol{p}_i\cdot\boldsymbol{S}_j}{m_im_j}\Bigg),\\
        {V}_{ij}^{so(s)}&=-\frac{1}{r_{ij}}\frac{d V_{ij}^{conf}}{d r_{ij}}\Bigg(\frac{\boldsymbol{r}_{ij}\times\boldsymbol{p}_i\cdot\boldsymbol{S}_i}{2m_i^2}-\frac{\boldsymbol{r}_{ij}\times\boldsymbol{p}_j\cdot\boldsymbol{S}_j}{2m_j^2}\Bigg), \notag
    \end{split}
\end{align}
where $S_i$ and $\boldsymbol{p}_i$ are the spin and momentum operators of the $i$-th quark, respectively.

The parameters of the quark potential model adopted in this work have been presented in Tab. \ref{tab:potential parameters}. The confinement strength $b_{ss}$ and the strong coupling constant $\alpha_{ss}$ between two strange quarks are taken from Ref.~\cite{Liu:2019wdr}, where they were determined by fitting the $\Omega^-$ mass spectrum. The constituent quark mass of the $s$ quark and $u/d$ quarks are fixed within their typical ranges.

\begin{table}[htbp]
    \centering
    \caption{Quark model parameters used in this work.}
    \begin{tabular}{ccccc}
    \hline\hline
        $b_{ss}\ (\text{GeV}^2)$  &$\alpha_{ss}$  &$\sigma_{ss}\ (\text{GeV})$  &$m_s\ (\text{GeV})$  &$m_{u/d}\ (\text{GeV})$\\
        \hline
         0.11 &0.77 &1.6 &0.5 &0.3\\
         \hline
         $b_{us}\ (\text{GeV}^2)$  &$\alpha_{us}$  &$\sigma_{us}\ (\text{GeV})$  &$C_0\ (\text{GeV})$  \\
        \hline
        0.12 &0.79 &1.8 & $-0.619$ \\
        \hline\hline
    \end{tabular}       \label{tab:potential parameters}
\end{table}

To obtain the masses and wave functions of baryons, we adopt the variation principle to solve the
Schr\"{o}dinger equation with the non-relativistic Hamiltonian
\begin{align}
    \begin{split}
        \hat{H}|\Psi_{JJ_z}\rangle=E|\Psi_{JJ_z}\rangle,
    \end{split}
\end{align}
in which the baryon wave function $|\Psi_{J,J_z}\rangle$ is constructed as a combination of color ($\phi_c$), spin ($\chi_{SS_z}$), spatial ($\psi_{NLL_z}$), and flavor ($\phi_f$) terms:
\begin{align}
    \begin{split}
        \Psi_{JJ_z}&=\mathcal{A}\big\{[\psi_{NLL_z}^{\sigma}(\boldsymbol{\rho},\boldsymbol{\lambda})\otimes\chi^{\sigma}_{SS_z}]_{JJ_z}\phi_f\phi_c\big\},
    \end{split}
\end{align}
where $\mathcal{A}$ represents the operator that imposes the antisymmetry on the total wave function when the two light quarks are exchanged, and the spatial wave function $\psi_{NLL_z}^{\sigma}(\boldsymbol{\rho},\boldsymbol{\lambda})$ consists of both $\rho$-mode and $\lambda$-mode excitations. We use the H.O. wave function bases, i.e.,
\begin{align}
    \begin{split}
        \psi_{n_{\zeta}l_{\zeta}m_{\zeta}}(\alpha_{\zeta\ell},\zeta)=R_{n_\zeta l_{\zeta}}(\alpha_{\zeta\ell},\zeta)Y_{l_{\zeta}m_{\zeta}}(\hat{\zeta}),
    \end{split}
\end{align}
where $Y_{l_{\zeta}m_{\zeta}}(\hat{\zeta})$ ($\zeta=\rho,\lambda$) is the spherical harmonic function. The radial part $R_{n_\zeta l_{\zeta}}(\alpha_{\zeta},\zeta)$ in coordinate space can be written as
\begin{eqnarray}
        R_{n_{\zeta}l_{\zeta}}(\alpha_{\zeta\ell},\zeta) &=& \alpha_{\zeta\ell}^{\frac32}\left[\frac{2^{l_{\zeta}+2-n_{\zeta}}(2l_{\zeta}+2n_{\zeta}+1)!!}{\sqrt{\pi}n_{\zeta}![(2l_{\zeta}+1)!!]^2}\right]^{\frac12} \times  \nonumber \\
        && \!\!\! \!\!\!\!\!\! \!\!\!\!\!\!  (\alpha_{\zeta\ell}\zeta)^l e^{-\frac12\alpha_{\zeta\ell}^2\zeta^2}F(-n_{\zeta},l_{\zeta}+\frac32,\alpha_{\zeta\ell}^2\zeta^2),
\end{eqnarray}
where $F(-n_{\zeta},l_{\zeta}+\frac32,\alpha_{\zeta\ell}^2\zeta^2)$ is the confluent hypergeometric function. Using the Fourier transformation, the form of the radial part in momentum space is
\begin{widetext}
\begin{eqnarray}
&& R_{n_{\zeta}l_{\zeta}}(\alpha_{\zeta\ell},p_\zeta) = (-1)^{n_{\zeta}}(-i)^{l_{\zeta}}\left[\frac{2^{l_{\zeta}+2-n_{\zeta}}(2l_{\zeta}+2n_{\zeta}+1)!!}{\sqrt{\pi}n_{\zeta}![(2l_{\zeta}+1)!!]^2}\right]^{\frac12}  \alpha_{\zeta\ell}^{-\frac32}(\frac{p_{\zeta}}{\alpha_{\zeta\ell}})^l e^{-\frac{p_{\zeta}^2}{2\alpha_{\zeta\ell}^2}}F(-n_{\zeta},l_{\zeta}+\frac32,\frac{p_{\zeta}^2}{\alpha_{\zeta\ell}^2}),
\end{eqnarray}
\end{widetext}
where $\alpha_{\zeta\ell}$ can be related to the H.O. frequency $\omega_{\zeta\ell}$ with $\alpha_{\zeta\ell}\equiv 1/d_{\zeta\ell}=\sqrt{M_{\zeta}\omega_{\zeta\ell}}$. The reduced masses $M_{\rho,\lambda}$ are defined by $M_{\rho}\equiv\frac{2m_1m_2}{m_1+m_2}$ and $M_{\lambda}\equiv \frac{3(m_1+m_2)m_3}{2(m_1+m_2+m_3)}$. For a system with three identical quark [e.g., $\Omega^-(sss)$], one has $\alpha_{\lambda\ell}=\alpha_{\rho\ell}$. However, for systems consisting of two identical quarks and another quark being different [e.g., $\Xi^0(ssu)$], the parameters of the two modes are not independent and their relationship is $\alpha_{\lambda\ell}=\alpha_{\rho\ell}(\frac{3m^{\prime}}{2m+m^{\prime}})^{1/4}$. Taking into account that the constituent mass of a strange quark is
close to that of a $u/d$ quark the relation $d_{\lambda\ell}\simeq d_{\rho\ell}$ is a good approximation~\cite{Xiao:2013xi}. Thus, the spatial wave function can be expanded as
\begin{align}
    \psi_{NLL_z}^{\sigma}(\boldsymbol{\rho},\boldsymbol{\lambda})=\sum_{\ell}^nC_{\ell}\psi_{NLL_z}^{\sigma}(d_{\ell},\boldsymbol{\rho},\boldsymbol{\lambda}).
\end{align}

Then, the Schr\"{o}dinger equation can be solved by dealing with the generalized eigenvalue problem,
\begin{align}
    \begin{split}
        \sum_{\ell^{\prime}=1}^{n}(H_{\ell\ell^{\prime}}-EN_{\ell\ell^{\prime}})C_{\ell^{\prime}}=0,
    \end{split}
\end{align}
where $H_{\ell\ell^{\prime}}\equiv \langle\xi(d_{\ell}^{\prime})|H|\xi(d_{\ell})$ and $N_{\ell\ell^{\prime}}\equiv \langle\xi(d_{\ell}^{\prime})|\xi(d_{\ell})$, while the function $\xi(d_{\ell})$ is given by \begin{align}
    \begin{split}
        \xi(d_{\ell})=\sum_{L_z+S_z=J_z}\langle LL_z;SS_z|JJ_z\rangle\psi_{NLL_z}^{\sigma}(d_{\ell},\boldsymbol{\rho},\boldsymbol{\lambda})\chi_{SS_z}^{\sigma}. \notag
    \end{split}
\end{align} 

In the calculations, the variational parameter $d_{\ell}$ is selected to form a geometric progression~\cite{Hiyama:2003cu},
\begin{align}
    d_{\ell}=d_1a^{\ell-1}(\ell=1,\cdots n),
\end{align}
where $n$ represents the number of basis functions, and $a$ is the ratio coefficient. There are three parameters $\{d_1,d_n,n\}$ to be determined with the variation method. It is found that taking $d_1=0.1$ fm, $d_n=2$ fm, and $n=10$, we can obtain stable results for the $\Xi$ baryons.

\begin{table*}[htbp]
    \centering
    \caption{The obtained masses (MeV) of the $\Xi$ states compared with the experimental data~\cite{ParticleDataGroup:2024cfk} (labeled with Expt.) and other theoretical calculations~\cite{Capstick:1986ter, Menapara:2021dzi, Faustov:2015eba, Loring:2001ky, Bijker:2000gq, Oh:2007cr, Pervin:2007wa, Chen:2009de, Engel:2013ig}. The values outside and inside the parentheses in the fourth column are the resonance masses without and with first order correction respectively in Ref~\cite{Menapara:2021dzi}. The harmonic oscillator (H.O.) strength parameters $\alpha_{\rho}$ (GeV), given in the last column, are extracted by fitting a single H.O. form to our numerical wave functions, with the constraint that the root-mean-square radius of the $\rho$-mode excitations be reproduced. }
    \begin{tabular}{cccccccccccccc}
    \hline\hline
         $n^{2S+1}L_{J^P}$  & Ours  &Ref.~\cite{Capstick:1986ter} &Ref.~\cite{Menapara:2021dzi} &Ref.~\cite{Faustov:2015eba} &Ref.~\cite{Loring:2001ky} &Ref.~\cite{Bijker:2000gq} &Ref.~\cite{Oh:2007cr} &Ref.~\cite{Pervin:2007wa} &Ref.~\cite{Chen:2009de} &Ref.~\cite{Engel:2013ig} &Expt.~\cite{ParticleDataGroup:2024cfk} &$\alpha_{\rho}$  \\
         \hline
         $1^2S_{1/2^+}$ &$1319.2$ &1305 &1322(1321) &1330 &1310 &1334 &1318 &1325 &1317 &$1303\pm13$ &1318 &0.460\\
         $1^4S_{3/2^+}$ &$1530.0$ &1505 &1531(1524) &1518 &1539 &1524 &1539 &1520 &1526 &$1553\pm18$ &1530 &0.373\\
         $1^2P_{1/2^-}$ &$1732.9$ &1755 &1886(1889) &1682 &1770 &1869 &1658 &1725 &1772 &$1716\pm43$ &1690$(?)$ &0.463\\
         $1^2P_{3/2^-}$ &$1820.7$ &1785 &1871(1873) &1764 &1780 &1828 &1820 &1759 &1801 &$1906\pm 29$ &1820 &0.404\\
         \hline\hline
    \end{tabular}     \label{tab:Xi mass}
\end{table*}

The model parameters $b_{us}$, $\alpha_{us}$, $\sigma_{ss/us}$ and $C_0$ in Tab.~\ref{tab:potential parameters} are determined by an overall description of the masses of the four $\Xi$ states~\cite{ParticleDataGroup:2024cfk}: 4-star ground state $\Xi(1318)$, 4-star sate $\Xi(1530)$, the 3-star $\Xi(1690)$ and $\Xi(1820)$ resonances. Although there is still controversy over the nature and even the $J^P$ assignment of $\Xi(1690)$, as mentioned in Sec.~\ref{sec-1 to 3}, some experimental analyses confirm its quantum numbers of $J^P=1/2^-$. In addition, theoretical analysis based on the quark model suggests that it might be the first orbital excitation state $1^2P_{1/2^-}$, although some other calculations predict the $1^2P_{1/2^-}$ mass to be slightly higher than the experimental $\Xi(1690)$ mass. The obtained mass spectra for these four $\Xi$ states compared with the data and some other works are given in Tab.~\ref{tab:Xi mass}. We compare the obtained first orbital excitation state $1^2P_{1/2^-}$ with $\Xi(1690)$. But, its obtained theoretical mass is higher than the experimental mass of $\Xi(1690)$. {Nevertheless, we do not claim that the obtained $1^2P_{1/2^-}$ state is the $\Xi(1690)$. This is why we placed a question mark (?) next to $\Xi(1690)$ in Tab.~\ref{tab:Xi mass}. In the following calculations, we adopt a mass of  1732.9 MeV for the $1^2P_{1/2^-}$ state, rather than the experimental mass of $\Xi(1690)$.}

It is worth noting that the purpose of this work is not to determine the mass spectrum of $\Xi$ baryons. Instead, we fit several experimentally well-established baryons to constrain the parameters of the quark potential model, which in turn allows us to obtain numerical wave functions for the $\Xi$ states. The stability and rationality of these numerical wave functions are evaluated by comparing the obtained mass from solving the Schr\"{o}dinger equation with the corresponding experimental values. The resulting numerical wave functions will be adopted in subsequent calculation of weak decays within the NRCQM, thereby reducing uncertainties introduced by additional phenomenological parameters.

\section{The hadron-level amplitudes and Final state interactions} \label{sec-hardon-level amp}

Neglecting differences from final-state hadronization, the DPE amplitude $\mathcal{M}_{\text{DPE}}(\Omega^-\to\Xi^0\pi^-)$ for $\Omega^-\to\Xi^0\pi^-$ and the CS amplitude $\mathcal{M}_{\text{CS}}(\Omega^-\to\Xi^-\pi^0)$ for $\Omega^-\to\Xi^-\pi^0$ satisfy the relation
\begin{align}
    \begin{split}
        \Bigg|\frac{\mathcal{M}_{\text{CS}}(\Omega^-\to\Xi^-\pi^0)}{\mathcal{M}_{\text{DPE}}(\Omega^-\to\Xi^0\pi^-)}\Bigg|=\frac{1}{3\sqrt{2}},
        \label{eq:amp rate}
    \end{split}
\end{align}
 where $1/3$ and $1/\sqrt{2}$ are respectively the color suppression factor and the isospin factor of $\pi^0$ (see Fig.~\ref{fig:quark-level diagram}). Then taking into account the phase spaces for these two channels are nearly identical, the relation of the decay branching ratios ($BR$) contributed by the $s\to ud\bar{u}$ weak transitions for these two channels is $BR_{\text{DPE}}(\Omega^-\to \Xi^0\pi^-)\simeq 18\times BR_{\text{CS}}(\Omega^-\to \Xi^-\pi^0)$. {This ratio is apparently different from the experimental value of $BR_{\text{exp}}(\Omega^-\to \Xi^0\pi^-)\simeq 3\times BR_{\text{exp}}(\Omega^-\to \Xi^-\pi^0)$ which was measured by the BESIII Collaboration~\cite{BESIII:2023ldd}. Moreover, as shown in Fig.~\ref{fig:quark-level diagram} (a) and (c), the transition of  $\Omega^- \to \Xi^0\pi^-$ via the DPE implies a larger $BR$ than that of $\Omega^- \to \Lambda K^-$ via the CS transition. However, this implication is inconsistent with experimental data, which show that the decay of $\Omega^-\to\Lambda K^-$ has the  largest branching ratio among the decays involving the $s\to ud\bar{u}$ transition~\cite{BESIII:2023ldd}. Such a phenomenon suggests that the $s\to ud\bar{u}$ transition at the tree level is insufficient to explain the $\Omega^-$ hadronic weak decays, and additional mechanisms should be considered in the transition amplitude. Thus, the contribution of the pole terms that only appear in $\Lambda K^-$ channel cannot be ignored.}

\begin{figure}[htbp]
    \centering
    \includegraphics[scale=2.2]{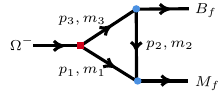}
     \caption{Schematic diagram of the FSIs from the intermediate channel rescattering to $\Xi^-\pi^0$. Black squares and red dots represent weak and strong vertices respectively.}
    \label{fig:loop}
\end{figure}

To account for the discrepancy between the expected ratio of 1/3 and the obtained value of 1/18, it is natural to consider possible enhancements to $\Omega^-\to \Xi^-\pi^0$ from the FSIs. Notably, both $\Omega^-\to \Xi^0\pi^-$ and $\Omega^-\to \Lambda K^-$ exhibit relatively large branching fractions, which is consistent with the dominant DPE process in the former and a sizable pole term contribution in the latter. As a result, the leading FSI correction to $\Omega^-\to \Xi^-\pi^0$ is expected to arise primarily from rescattering via the intermediate channels $\Xi^0\pi^-$ and $\Lambda K^-$, as shown in Fig.~\ref{fig:loop} through the hadronic level triangle diagrams.

We employ the effective field theory approach to compute the loop-level amplitudes. The explicit effective interaction Lagrangians for the above vertices can be written as~\cite{Janssen:1996kx, Riska:2000gd}
\begin{eqnarray}
        \mathcal{L}_{DBP} &=& ig_{DBP}^{(\text{PC})}\bar{D}^{\mu}\partial_{\mu}PB\\
        \mathcal{L}_{BBP} &=& i\frac{f_{BBP}}{m_{P}}\bar{B}\gamma_{\mu}\gamma_5B\partial^{\mu}P,\\
        \mathcal{L}_{BBV} &=& -g_{BBV}\bar{B}\gamma_{\mu}BV^{\mu}+\frac{f_{BBV}}{2m_B}\bar{B}\sigma^{\mu\nu}B\partial_{\nu}V_{\mu},\\
        \mathcal{L}_{VPP} &=& ig_{VPP}\langle(P\partial_{\mu}P-\partial_{\mu}PP)V_{\mu}\rangle,
        \label{eq:Lagrangian}
  \end{eqnarray}
where $\sigma^{\mu\nu}=i[\gamma^{\mu},\gamma^{\nu}]/2$. ${D}$ and ${B}$ represent decuplet baryon field with $J^P=3/2^+$ and octet baryon field with $J^P=1/2^+$, respectively; $P$ and $V$ represent pseudo-scalar meson with $J^P=0^-$ and vector meson with $J^P=1^-$, respectively. 

Notation $[M1,M3,(M2)]$ is used to denote the intermediate interaction between particle $M1$ and $M3$ by exchanging $M2$. The masses and 4-vector momenta of these internal particles are denoted by $(m_1,m_2,m_3)$ and $(p_1,p_2,p_3)$, respectively. The 4-vector momenta of the initial-state baryon, final-state baryon and final-state meson are labeled as $p_i$, $p_f$ and $p_{M}$, respectively. In order to cut off the ultra-violet (UV) divergence in the loop integrals, we include a commonly-adopted form
factor to regularize the loop integration:
\begin{align}
    \mathcal{F}(p_i^2)=\prod_i(\frac{\Lambda_i^2-m_i^2}{\Lambda_i^2-p_i^2})
\end{align}
where $\Lambda_i$ is the cut-off parameter. {This UV cut-off scheme has been widely applied in processes such as scattering and photoproduction~\cite{Holzwarth:1996bc,Gross:1991pm,Sato:1996gk,Zhao:1998fn,Zhao:2001ue}. The value of $\Lambda$ depends strongly on the specific reaction process and is typically determined by fitting experimental data generally. For instance, the typical value of $\Lambda$ in meson photoproduction is around 1 GeV~\cite{Oh:2001bq,Zhao:1998fn,Zhao:2001ue}; in the Born potential of $\pi N$ scattering, $\Lambda$ is taken as 1.3 GeV~\cite{Liu:1995st}; and in the one-boson-exchange model for $NN$ scattering, $\Lambda$ ranges from 1.3 to 2 GeV for different intermediate mesons~\cite{Machleidt:1987hj}. In this work, we adopt a unified cutoff parameter $\Lambda$ to minimize the number of free parameters and determine its value based on experimental data.} 

As follow, we will write down the detailed amplitudes for each loop transition.
\begin{itemize}
    \item $[\pi^-,\Xi^0,(\rho^+)]$ and $[K^-,\Lambda,(K^{*+})]$
    \begin{widetext}  
    \begin{align}
        \begin{split}
            &i\mathcal{M}=g_1g_2g_3\int\frac{d^4p_1}{(2\pi)^4}\frac{p_{1\mu}(p_1+p_{M})_{\nu}(g^{\alpha\nu}-\frac{p_{2}^{\alpha}p_2^{\nu}}{p_2^2})\bar{u}(p_{f},\lambda_{f})(\gamma_{\alpha}-\frac{f_3}{2m_{B}g_3}\frac{[\gamma_{\alpha},\gamma_{\beta}]}{2}p_2^{\beta})(\slashed{p_3}+m_3)u^{\mu}(p_{i},\lambda_{i})}{(p_1^2-m_1^2+i\epsilon)(p_2^2-m_2^2+i\epsilon)(p_3^2-m_3^2+i\epsilon)}\mathcal{F}(p_i^2).
        \end{split}
    \end{align}
     \end{widetext}

    \item $[\Xi^0,\pi^-,(\bar{\Xi}^0)]$ and $[\Lambda,K^-,(\bar{\Sigma}^0)]$
    
    \begin{widetext}
     \begin{align}
            &i\mathcal{M}=g_1g_2g_3\int\frac{d^4p_1}{(2\pi)^4}\frac{p_{3\mu}p_3^{\alpha}p_M^{\nu}\bar{u}(p_f,\lambda_f)\gamma_{\alpha}\gamma_5(\slashed{p}_2+m_2)\gamma_{\nu}\gamma_5(\slashed{p}_1+m_1)u^{\mu}(p_i,\lambda_i)}{(p_1^2-m_1^2+i\epsilon)(p_3^2-m_3^2+i\epsilon)(p_2^2-m_2^2+i\epsilon)}\mathcal{F}(p_i^2).
    \end{align}
     \end{widetext}
   
\end{itemize}
In these above expressions, we do not distinguish the coupling constants at the hadronic vertices, but just denote them as $g_i$ ($i=1,2,3$) for conciseness. $u^{\mu}(p,\lambda)$ represents the Rarita–Schwinger spinor of a spin-3/2 baryon with momentum $p$ and helicity $\lambda$. It can be constructed from the Dirac spinor $u(p,\lambda)$ and the polarization vector $\epsilon^{\mu}(p,\lambda)$,
\begin{align}
    \begin{split}
        u^{\mu}({p},\lambda)&=\sum_{\lambda_1,\lambda_2}\langle 1\lambda_1;\frac12\lambda_2|\frac32\lambda\rangle\epsilon^{\mu}({p},\lambda_1)u({p},\lambda_2).
    \end{split}
\end{align}

Next, we explain how these coupling constants of each vertex in the triangle loop shown in Fig.~\ref{fig:loop} are determined. Previous calculations indicate that the PV transition amplitude vanishes for $s\to ud\bar{u}$ processes, and that the amplitude of $su\to ud$ processes is significantly suppressed compared to the PC one. Therefore, both the amplitude of $s\to ud\bar{u}$ and the amplitude of $su\to ud$ are dominated by the PC transition, and we only consider the loop diagrams of the PC transition. The weak couplings $g_{\Omega\Xi^0\pi^-}^{(\text{PC})}=1.37\times 10^{-6}\ \text{GeV}^{-1}$ and $g_{\Omega\Lambda K^-}^{(\text{PC})}=3.98\times 10^{-6}\ \text{GeV}^{-1}$ are extracted from the tree processes illustrated in Fig.~\ref{fig:quark-level diagram} within the NRCQM. 

The coupling constants of the $BBP$ and $BBV$ vertices can be determined by the SU(3) relationship,
\begin{align}
    \begin{split}
        &f_{NN\pi}=f_8^P,\ \ \ f_{\Xi\Xi\pi}=(1-2\alpha_{BBP})f_8^P,\\
        &f_{\Lambda\Xi K}=\frac{1}{\sqrt{3}}(4\alpha_{BBP}-1)f_8^P,\\
        &f_{\Lambda\Sigma\pi}=\frac{2}{\sqrt{3}}(1-\alpha_{BBP})f_8^P\\
         &g_{NN\rho}=f_8^V,\ \ \ g_{\Xi\Xi\rho}=(1-2\alpha_{BBV}^e)g_8^V,\\
         &g_{\Lambda\Xi K^*}=\frac{1}{\sqrt{3}}(4\alpha_{BBV}^e-1)g_8^V,
    \end{split}
\end{align}
where $f_8^P=f_{NN\pi}=0.989$~\cite{Janssen:1996kx} and $\alpha_{BBP}=0.365$~\cite{Rijken:1998yy} are used. For the vector coupling constant, we take $g_{8}^V=g_{NN\rho}=3.25$ from the dispersion theory based on the meson exchange model~\cite{Janssen:1996kx}. The ratio $\alpha_{BBV}^e$ is determined by the relationship between the $NN\omega$ coupling $g_{NN\omega}=15.83$ based on the Bonn model~\cite{Machleidt:1987hj} and the coupling $g_{NN\rho}$: $\alpha_{BBV}^e=\frac14\times(\frac{g_{NN\omega}}{g_{NN\rho}}+1)=1.47$, which is different from unity required by the quark model. As for the tensor coupling constant $f_{BBV}$, the Nijmegen model gives $f_{NN\rho}=12.52$~\cite{Rijken:1998yy} and the corresponding ratio $\alpha_{BBV}^m$ is usually fixed at 0.4~\cite{Pais:1966eox,Sakita:1965qt}, which is obtained by the SU(6) symmetry. For the $VPP$ vertices, the couplings $g_{\rho^+\pi^-\pi^0}=\sqrt{2}g_{VPP}$ and $g_{K^{*+}K^-\pi^0}={g_{VPP}}/{\sqrt{2}}$ are extracted from the experimental data in $\rho\to \pi\pi$ and $K^*\to K\pi$ decays~\cite{Cheng:2021nal, Cao:2023gfv}. These strong coupling constants of the $BBP$, $BBV$ and $VPP$ vertices used in this work are summarized in Tab.~\ref{tab:BBP, BBV and VPP}. 

\begin{table}[htbp]
    \centering
    \caption{The strong coupling constants of the $BBP$, $BBV$ and $VPP$ vertices.}
    \label{tab:BBP, BBV and VPP}
    \begin{tabular}{c|cccc}
    \hline\hline
         $BBP$&$f_{\Xi\Xi\pi}$  &$f_{\Sigma\Xi K}$ &$f_{\Lambda\Sigma \pi}$\\
         \hline
         Values &0.267&0.989&0.725\\
         \hline
         $BBV$&$g_{\Xi\Xi\rho}$ &$g_{\Lambda\Xi K^*}$&$f_{\Xi\Xi\rho}$ &$f_{\Lambda\Xi K^*}$\\
         \hline
         Values &$-6.31$&$9.16$&$10.86$&$-1.28$\\
         \hline
         $VPP$ &$g_{\rho^+\pi^0\pi^-}$ &$g_{K^{*+}K^-\pi^0}$\\
         \hline
         Values &$5.96$ &$3.30$ \\
         \hline\hline
    \end{tabular}
\end{table}

\section{The decay width and asymmetry parameter}

The decay amplitudes of the DPE, CS, and pole terms are calculated within the framework of the NRCQM, with mesons and baryons represented by mock states~\cite{Hayne:1981zy}:
\begin{eqnarray}
&& |\mathbb{M}(\boldsymbol{ P}_c;J,J_z)\rangle = \sum_{S_z,L_z;c_i}\langle L L_z;S S_z|J J_z\rangle\int d\boldsymbol{p}_1d\boldsymbol{p}_2 \nonumber \\
        && \times \delta^3(\boldsymbol{p}_1+\boldsymbol{p}_2-\boldsymbol {P}_c) \psi_{NLL_z}(\boldsymbol{p}_1,\boldsymbol{p}_2) \frac{\chi_{S,S_z}^{s_1,s_2} \delta_{c_1c_2} \phi_{i_1i_2}}{\sqrt{3}} \nonumber \\
        && \times|q_1(\boldsymbol{p}_1)\bar{q}_2(\boldsymbol{p}_2)\rangle, \\
&& |\mathbb{B}(\boldsymbol{P}_c;J,J_z)\rangle = \sum_{S_z,L_z;c_i}\langle L L_z;S S_z|J J_z\rangle\int d\boldsymbol{p}_1d\boldsymbol{p}_2d\boldsymbol{p}_3  \nonumber \\ 
        && \times \delta^3(\boldsymbol{p}_1+\boldsymbol{p}_2+\boldsymbol{p}_3-\boldsymbol{P}_c) \psi_{NLL_z}(\boldsymbol{p}_1,\boldsymbol{p}_2,\boldsymbol{p}_3) \chi_{S,S_z}^{s_1,s_2,s_3} \nonumber \\
        && \times\frac{\epsilon_{c_1c_2c_3}}{\sqrt{6}}\phi_{i_1i_2i_3}|q_1(\boldsymbol{p}_1)q_2(\boldsymbol{p}_2)q_3(\boldsymbol{p}_3)\rangle,
        \label{Eq: mock states}
\end{eqnarray}
where $c_j$, $s_j$, $i_j$ are color, spin, and flavor indexes, respectively. $\psi_{NLL_z}$ is the spatial wave function which is taken as an harmonic oscillator wave function. $\boldsymbol{p}_i$ denotes the single quark (antiquark) three-vector momentum, and $\boldsymbol{P}_c$ ($\boldsymbol{P}_c^{\prime}$) denotes the hadron momentum. $\chi_{S,S_z}$ is the spin wave function; $\phi$ is the flavor wave function, and $\delta_{c_1c_2}/\sqrt{3}$ and $\epsilon_{c_1c_2c_3}/\sqrt{6}$ are the color wave functions for the meson and baryon, respectively. The normalization condition for the mock states are:
\begin{align}
    \begin{split}
        \langle \mathbb{M}(\boldsymbol{P}_c^{\prime};J,J_z)|\mathbb{M}(\boldsymbol{P}_c;J,J_z)\rangle&=\delta^3(\boldsymbol{P}_c^{\prime}-\boldsymbol{P}_c),\\
        \langle \mathbb{B}(\boldsymbol{P}_c^{\prime};J,J_z)|\mathbb{B}(\boldsymbol{P}_c;J,J_z)\rangle&=\delta^3(\boldsymbol{P}_c^{\prime}-\boldsymbol{P}_c).
        \label{Eq: normalization condition for the mock states}
    \end{split}
\end{align}

In the above convention, for the two-body decay $A \to B + C$, the $S$-matrix is defined by
\begin{align}  
    \begin{split}  
        \mathcal{S}=\mathcal{I}-2\pi i\delta^4(P_A-P_B-P_C)\mathcal{\tilde{M}},  
    \end{split}  
\end{align} 
where 
\begin{align}  
    \begin{split}  
        \delta^3(\boldsymbol{P}_A-\boldsymbol{P}_B-\boldsymbol{P}_c)\mathcal{\tilde{M}}\equiv\langle BC|H_I|A\rangle.  
    \end{split}  
\end{align} 
By integrating over phase space, the decay width can be expressed as~\cite{Niu:2020gjw}:
\begin{align}  
    \begin{split}  
        \Gamma(A\to B+C)&=\frac{8\pi^2|\boldsymbol{k|}E_{{B}}E_{{C}}}{M_A}\frac{1}{2J_A+1}\sum_{\text{spin}}|\mathcal{\tilde{M}}|^2, 
        \label{eq:Gamma-1}
    \end{split}  
\end{align}
where $\boldsymbol{k}$ is the three-vector momentum of the final-state meson in the initial state rest frame, and $J_A$ is the spin of the initial state. $E_B$ and $E_C$ are the energies of the final states $B$ and $C$, respectively. 

By redefining 
\begin{equation}
    \mathcal{{M}}\equiv 8\pi^{3/2} (M_AE_BE_C)^{1/2}\mathcal{\tilde{M}} \ ,
\end{equation}
where $\mathcal{{M}}$ is the transition matrix element defined at the hadronic level, we can unify all the transition amplitudes as follows:
\begin{align}
    \begin{split}
        \mathcal{S}&=1+i\mathcal{T}=1+(2\pi)^4i\delta^4(P_A-P_B-P_C)\mathcal{M} \ .
    \end{split}
\end{align}
In this convention, the expression for the partial decay width is then given by~\cite{Li:1996yn, Cao:2025kvs}:
\begin{align}
    \begin{split}
        &\Gamma(A\to B+C)= \frac{|\boldsymbol{k|}}{8\pi M_A^2}\frac{1}{2J_A+1} \times \\
&\sum_{J_i^z,J_f^z}\Big(\Big|\mathcal{M}^{J_f,J_f^z;J_i,J_i^z}_{(\text{PC})}\Big|^2+\Big|\mathcal{M}^{J_f,J_f^z;J_i,J_i^z}_{(\text{PV})}\Big|^2\Big),
        \label{eq:Gamma-2}
    \end{split}
\end{align}
where the non-vanishing helicity amplitudes have the following relationship:
\begin{align}
    \begin{split}
        &\mathcal{M}_{(\text{PC})}^{\frac12,+\frac12;\frac32,+\frac12}=-\mathcal{M}_{(\text{PC})}^{\frac12,-\frac12;\frac32,-\frac12},\\
        &\mathcal{M}_{(\text{PV})}^{\frac12,+\frac12;\frac32,+\frac12}=+\mathcal{M}_{(\text{PV})}^{\frac12,-\frac12;\frac32,-\frac12}.
    \end{split}
\end{align}

We can also calculate the asymmetry parameters in our model. For the interest of understanding the transition mechanisms, we will focus on 
\begin{align}
    \begin{split}
        \alpha &=\frac{2\text{Re}(A^*B)}{|A|^2+|B|^2} \\
        & =\frac{2\text{Re}\Big[(\mathcal{M}_{(\text{PV})}^{\frac12,-\frac12;\frac32,-\frac12})^*\mathcal{M}_{(\text{PC})}^{\frac12,-\frac12;\frac32,-\frac12}\Big]}{|\mathcal{M}_{(\text{PV})}^{\frac12,-\frac12;\frac32,-\frac12}|^2+|\mathcal{M}_{(\text{PC})}^{\frac12,-\frac12;\frac32,-\frac12}|^2}, 
    \end{split}
\end{align}
where $A$ and $B$ represent the $D$-wave (PV) and $P$-wave (PC) amplitudes, respectively. It is obvious that $\alpha$ is bounded by $-1\leq\alpha\leq 1$ and characterizes the relative strength between the PC and PV transitions.

\section{Numerical results and discussions}\label{sec-numerical-results}

Before presenting the numerical results, we clarify the parameters and inputs in our calculation as follows: the constituent quark masses for the $u$, $d$, and $s$ quarks are taken as $m_u=m_d=300$ MeV and $m_s=500$ MeV, as listed in Tab.~\ref{tab:potential parameters}; the masses of the $\Xi$ baryon states are given by solving the Schr\"{o}dinger equation in Sec.~\ref{sec-baryon wave functions}. Correspondingly, the numerical wave functions are also determined. For simplicity, the H.O. strength parameters $\alpha_{\rho}$ can be obtained by fitting a single H.O. form to our numerical wave functions, with the constraint that the root-mean-square radius of the $\rho$-mode excitations be reproduced. And the resulting values are listed in the last column of Tab.~\ref{tab:Xi mass}. 

For the initial $\Omega^-$ baryon, the H.O. parameters $\alpha_{\rho}=\alpha_{\lambda}=0.455$ GeV can be obtained from Ref.~\cite{Liu:2019wdr}, and this value was also derived by using the numerical wave function given in this literature to fit the single H.O. wave function. For $\Omega^-\to \Lambda K^-$ decay, the $\rho$-mode H.O. parameter of the final baryon $\Lambda$ is taken as $\alpha_{\rho}=0.447$ GeV from Ref.~\cite{Cao:2025kvs}, while the $\lambda$-mode H.O. parameter $\alpha_{\lambda}$ in a system with non-identical quarks is related to the $\rho$-mode parameter by $\alpha_{\lambda}=\big(\frac{3m^{\prime}}{m+m+m^{\prime}}\big)^{1/4}\alpha_{\rho}$, where $m^{\prime}$ is the mass of the quark that differs from the other two, for example, in the case of $\Lambda$ baryon, $m^{\prime}=m_s$.

The pion wave function is also expressed as a H.O. form with a parameter $R$. Since the pion meson is extremely light and associated with the spontaneous chiral symmetry breaking, it favors a larger value for $R$, and turns out to have a broad distribution in the momentum space~\cite{Kokoski:1985is}. A larger value for $R$ also manifests that the relativistic effects become predominant in such a light quark system. In this work, $R=0.53$ GeV was adopted, and this treatment is empirical and inevitably carries some intrinsic uncertainty. However, we would like to investigate the uncertainty by varying the parameter $R$ within a reasonable range in the subsequent numerical results and discuss the phenomenological consequences. Furthermore, for the kaon meson, we adopted the same $R$ value as that of the pion meson as the initial input.

The intermediate states of the pole terms are $\Xi$ resonances in the decay $\Omega^-\to \Lambda K^-$, as illustrated in Fig.~\ref{fig:quark-level diagram} (d). The states with $J^P=1/2^+$ and $J^P=1/2^-$ contribute to the PC and PV transition processes, respectively. To be more specific, we consider the ground state $1^2S_{1/2^+}$ and the first orbital excitation state $1^2P_{1/2^-}$, whose masses and wave functions are determined from the quark potential model and listed in Tab.~\ref{tab:Xi mass}. {The width of the ground state $1^2S_{1/2^+}$ is negligible, whereas that of the excited $1^2P_{1/2^-}$ state is not. In our calculation, we explicitly incorporate its width, taking the central value of the experimentally observed $\Xi(1690)$ resonance. Although this might seem to introduce uncertainty, the subsequent amplitude results will show that its impact is very small.}

\begin{table}[htbp]
    \centering
    \caption{The tree-level amplitudes $\mathcal{M}_T^{\frac32-\frac12;\frac12-\frac12}$ ($10^{-7}$ GeV) with $J_f^z=J_i^z=-1/2$ for different channels.}
    \label{tab:tree amp}
    \scalebox{0.86}{
    \begin{tabular}{cccccc}
    \hline\hline
     &{DPE} &{CS-1} &{CS-2} &{PT} &{PT}\\
     &(PC) &(PC) &(PC) &(PC) &(PV)\\
    \hline
    $\Omega^-\to \Xi^0\pi^-$ &$9.70$ &- &- &- &-\\
    $\Omega^-\to \Xi^-\pi^0$ &- &$-2.24$ &- &- &-\\
    $\Omega^-\to \Lambda K^-$ &- &- &$1.43$ &$-18.69$ &$-0.114-0.002i$\\
    \hline\hline
    \end{tabular}}
\end{table}

We first calculate the tree-level amplitudes for the DPE, CS and PT processes, which are summarized in Tab.~\ref{tab:tree amp}. It should be noted that for all $s \to ud\bar{u}$ weak processes, the PV transition amplitudes vanish because the corresponding spin matrix elements are zero, hence, they are not included in the Tab.~\ref{tab:DPE and CS spin}. A notable feature is that the DPE process contributes exclusively to $\Omega^-\to \Xi^0\pi^-$. This exclusivity arises because the $d\bar{u}$ quark pair produced at the weak vertex can hadronize directly into the final-state $\pi^-$ meson [see Fig.~\ref{fig:quark-level diagram}(a)]. This direct hadronization mechanism implies no color suppression for the DPE amplitude, which consequently explains its larger magnitude compared to the CS amplitude.

\begin{table}[htbp]
    \centering
     \caption{Our results of branching ratio (BR) and asymmetry parameter (AP) for the decays $\Omega^-\to \Xi\pi$ and $\Lambda K^-$ compared with the experimental data~\cite{BESIII:2023ldd, ParticleDataGroup:2024cfk} (labeled with Expt.) and the other theoretical predictions~\cite{Carone:1991ni, Galic:1979hh, Borasoy:1999ip, Xu:2020jfr}. For $\Omega^-\to \Xi^-\pi^0$, the cut-off parameter $\Lambda=(1.92\pm 0.07)$ GeV is determined based on the experimental data.}
    \label{tab:BR and AP}
    \begin{tabular}{c|c|c|c} 
    \hline\hline
     & $\Omega^-\to\Xi^0\pi^-$  &$\Omega^-\to\Xi^-\pi^0$   &$\Omega^-\to\Lambda K^-$\\
     \hline
     \multicolumn{4}{c}{BR ($\%$)} \\
     Ours     &24.52  &$8.4 \pm 0.6$      &{65.83}\\
     Expt.~\cite{BESIII:2023ldd}  &$25.03 \pm 0.69$ &$8.43\pm 0.59$  & $66.3\pm 2.2$\\
     Ref.~\cite{Carone:1991ni} &17.18&8.59&78.94 \\
     Ref.~\cite{Borasoy:1999ip} &20.21&9.48&69.98 \\
     Ref.~\cite{Galic:1979hh} &$34.67$ &$5.36$ &$67.98$\\
     Ref.~\cite{Xu:2020jfr} &$23.6\pm 0.7$ &$8.6\pm 0.4$ &$67.8\pm 0.7$ \\
     \hline
      \multicolumn{4}{c}{AP ($\alpha$)} \\
      Ours   & 0    &0     & 0.0121 \\
     Expt.~\cite{ParticleDataGroup:2024cfk} &$0.09\pm 0.14$ &{$0.05\pm 0.21$} &$0.0154\pm 0.002$\\
     Ref.~\cite{Borasoy:1999ip}&0&0&$-0.015$\\
     \hline\hline
    \end{tabular}
\end{table}

Our calculations of the branching ratios and the asymmetry parameters are presented in Tab.~\ref{tab:BR and AP}. These results are compared with the experimental data and the predictions from other theoretical models. As in Tab.~\ref{tab:BR and AP}, the calculated branching ratio for $\Xi^0\pi^-$ agrees with experiment,  which necessarily means that the result for $\Xi^-\pi^0$ lies significantly below the measured value. In our calculation, the branching ratio of the $\Xi^-\pi^0$ channel contributed by the CS process as shown in Fig.~\ref{fig:quark-level diagram} is $1.29\%$, which is much smaller than the experimental measurement. The large difference in the branching ratios between $\Omega^-\to\Xi^-\pi^0$ and $\Xi^0\pi^-$ is mainly because of the absence of the DPE process in the former channel. However, the decay $\Omega^-\to \Lambda K^-$ also lacks a DPE contribution, yet its experimentally measured branching ratio is sizable, at approximately 66$\%$. Our analysis reveals that this channel receives a dominant contribution from pole terms. Specifically, only one type of pole diagram is relevant here, characterized by a strong interaction occurs first and then a weak interaction follows [see Fig.~\ref{fig:quark-level diagram}(d)]. In our calculation, the intermediate poles we considered are the $\Xi$ baryons of the $1^2S_{1/2^+}$ and $1^2P_{1/2^-}$ states, corresponding to the PC and PV transition, respectively. As shown in Tab.~\ref{tab:tree amp}, the amplitude from the pole terms is much lager than that of the CS in this channel, thereby enhancing its predicted branching ratio. To be more specific, the decay $\Omega^-\to \Lambda K^-$ is dominated by the pole process of the PC transition, as the PV amplitude is much smaller than the PC amplitude. This also indicates that it is reasonable to ignore the contribution from higher excitation states.

After including both the $s\to ud\bar{u}$ mechanism through $W$-emission (DPE and CS) and the $su\to ud$ mechanism via $W$-exchange (PT) at the quark level, we calculate the branching ratios and asymmetry parameters for the decays $\Omega^-\to \Xi\pi$ and $\Lambda K^-$. Our results show that the branching ratios for $\Omega^-\to \Xi^0\pi^-$ and $\Lambda K^-$ channels are in good agreement with the experimental data presented in Tab.~\ref{tab:BR and AP}, while a significant discrepancy persists for $\Omega^-\to \Xi^-\pi^0$ whose predicted branching ratio is only $1.29\%$--are below the measured value.

Moreover, the asymmetry parameters for both $\Omega^-\to\Xi^0\pi^-$ and $\Xi^-\pi^0$ are predicted to be zero, due to the vanishing PV amplitudes in their DPE or CS processes. In $\Omega^- \to \Lambda K^-$, however, the pole terms mechanism introduce a small non-zero PV amplitude, resulting in a very small but non-zero asymmetry parameter. All three decays are found to be dominated by parity-conserving ($P$-wave) transitions, which is consistent with experimental measurements that report asymmetry parameters very close to zero. Ref.~\cite{Jenkins:1991bt} also expects the $D$-wave component (related to PV transition) to be very small. Notably, for the $\Omega^- \to \Xi^0\pi^-$ and $\Xi^-\pi^0$ channels, the experimental uncertainties of asymmetry parameters are larger than the central values themselves, meaning our predictions lie well within the allowed experimental ranges.

We consider the FSI contributions through the triangle process shown in Fig.~\ref{fig:loop}, while neglecting the two-point bubble loop mentioned in Ref.~\cite{Cao:2025kvs}, as our results shows that their contributions to be negligible. Since the PV amplitudes in the DPE and pole process are very small or even zero, we consider only FSIs in the PC transition process. Treating FSIs as an additional mechanism that can significantly enhance $\Omega^-\to \Xi^-\pi^0$, we present the calculated  helicity amplitudes of each hadronic loop in Tab.~\ref{tab:loop amp} with the cut-off parameter $\Lambda=1.5$ GeV and 2 GeV adopted. The results in Tab.~\ref{tab:loop amp} show that the first two loop amplitudes are at the same order of magnitude as the tree-level CS amplitude listed in Tab.~\ref{tab:tree amp} for $\Omega^-\to \Xi^-\pi^0$.

\begin{table}[htbp]
    \centering
    \caption{The loop amplitudes $\mathcal{M}_L^{\frac32-\frac12;\frac12-\frac12}$ ($10^{-8}$ GeV) with $J_f^z=J_i^z=-1/2$ for $\Omega^-\to \Xi^0\pi^-$ at $\Lambda = 1.5$ and $2.0$ GeV.}
    \label{tab:loop amp}
    \scalebox{0.85}{
    \begin{tabular}{ccccc}
    \hline\hline
    $\Lambda$ (GeV) &{$[\pi^-,\Xi^0,(\rho^+)]$} &{$[K^-,\Lambda,(K^{*+})]$} &{$[\Xi^0,\pi^-,(\bar{\Xi}^0)]$} &{$[\Lambda,K^-,(\bar{\Sigma}^0)]$}  \\
    \hline
    $1.5$ &$9.25-6.14i$ &$-5.32-11.89i$ &$0.008-0.03i$ &$0.026-0.38i$\\
    $2.0$ &$10.68-18.92i$ &$-6.74-16.84i$ &$0.006-0.14i$ &$0.020-1.00i$\\
    \hline\hline
    \end{tabular}}
\end{table}

\begin{figure}
    \centering
    \includegraphics[width=0.95\linewidth]{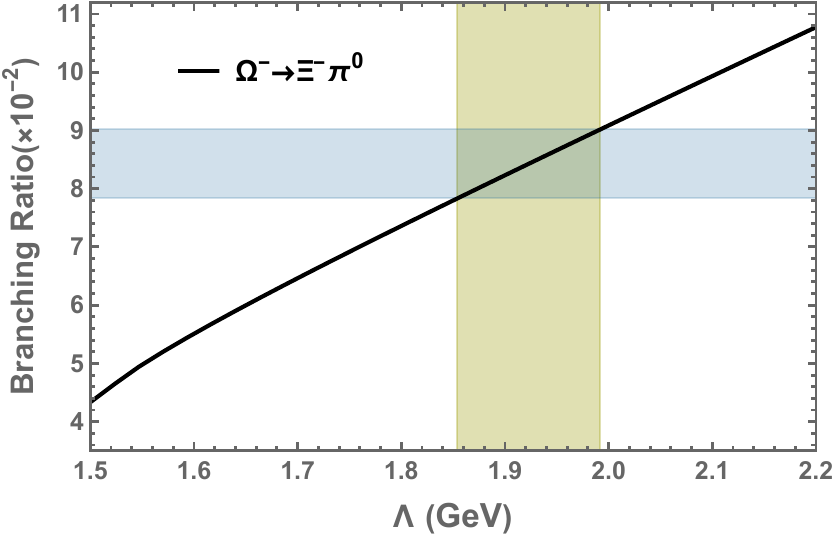}
    \caption{The obtained branching ratio of $\Omega^-\to \Xi^-\pi^0$ as a function of the cut-off parameter $\Lambda$, comparing with the experimental data (horizontal blue bands). The vertical dark yellow band indicates the range where our results overlap with the experimental data.}
    \label{fig:BR}
\end{figure}

By incorporating these FSIs, we plot the branching ratio of $\Omega^-\to \Xi^-\pi^0$ as a function of $\Lambda$, finding that for $\Lambda \simeq (1.92\pm 0.07)$ GeV, the predicted branching ratio agrees with the experimental value range~\cite{BESIII:2023ldd}, as shown in Fig.~\ref{fig:BR}. This consistency supports the interpretation that FSIs constitute the essential missing ingredient in a unified theoretical description of the $\Omega^-\to \Xi^-\pi^0$ decay. We look forward to BESIII collaboration making more precise measurements of the asymmetric parameters. We also examine the uncertainty of the results introduced by the H.O. parameter $R$. The $\Xi^0\pi^-$ and $\Lambda K^-$ channels are respectively dominated by the DPE process and the PT process. In addition, by including the FSI contributions through the rescattering processes $\Xi^0\pi^-/\Lambda K^-\to \Xi^-\pi^0$, the branching ratio of the $\Xi^-\pi^0$ channel is significantly enhanced. Consequently, the results of $\Xi^-\pi^0$ and $\Lambda K^-$ channels show that the $BR$ is highly insensitive on the H.O. parameter $R$, whereas the $\Xi^0\pi^-$ result is more sensitive to it. We observe that when $R=(0.53\pm 0.053)$ GeV, the  resulting branching ratio uncertainty for $\Xi^0\pi^-$ is $(24.52\pm 7.35)\%$. This level of uncertainty is reasonable and acceptable within the typical parameter dependence of the quark-model calculations. {Furthermore, since the $\Lambda K^-$ channel is dominated by the pole term and its theoretical prediction is highly stable against variations in $R$, we do not quote the corresponding uncertainty.}

\section{Summary} \label{sec-summary}

In this work, we have performed a systematic study of the singly Cabibbo-suppressed hadronic weak decays of the $\Omega^-$ hyperon, namely $\Omega^-\to \Xi^0\pi^-$, $\Xi^-\pi^0$, and $\Lambda K^-$. Within the NRCQM framework, we compute the tree-level decay amplitudes corresponding to the DPE, CS, and PT mechanisms originating from the quark-level transitions $s\to ud\bar{u}$ and $su\to ud$. For this calculation, we employ baryon numerical wave functions whose harmonic-oscillator parameters are fixed using available $\Xi$ states spectrum. This treatment effectively suppresses the uncertainties on predicted observables stemming from phenomenological parameter inputs.

The DPE process, due to the absence of color suppression, clearly plays a dominant role. This results in the branching ratio of the $\Xi^0\pi^-$ channel being much larger than that of the $\Xi^-\pi^0$. Therefore, when the branching ratio of $\Xi^0\pi^-$ is consistent with the experiment, $\Xi^-\pi^0$ inevitably deviates significantly from the experimental data. On the other hand, the large branching fraction of $\Omega^-\to \Lambda K^-$ is successfully reproduced by the dominant contribution from the pole terms involving intermediate $\Xi$ resonances ($1^2S_{1/2^+}$ and $1^2P_{1/2^-}$ states). 

Considering both the $s\to ud\bar{u}$ ($W$-emission) and $su\to ud$ ($W$-exchange) weak mechanisms, our results show that they cannot simultaneously describe the experimental branching ratios of these three channels. The main discrepancy lies in the $\Xi^-\pi^-$ channel, whose predicted branching ratio is significantly underestimated due to the much smaller amplitude of the CS process compared to the dominant DPE and pole term amplitudes. We find the deficit in the predicted branching ratio for $\Omega^-\to \Xi^-\pi^0$ can be remedied by incorporating the FSIs which occurs in the form of rescattering process. Specifically, we computed the loop amplitudes via triangle diagrams from the $\Xi^0\pi^-$ and $\Lambda K^-$ intermediate states into the $\Xi^-\pi^0$ final state. With the inclusion of the FSI contributions, our theoretical predictions for the branching ratios and asymmetry parameters of all three decay channels achieve excellent agreement with the latest experimental data~\cite{BESIII:2023ldd}. This demonstrates that a consistent theoretical description of $\Omega^-$ decays requires a comprehensive picture that integrates quark-level weak vertices, baryon pole structures, and hadron-level final-state rescattering effects. Our analysis highlights the significant role played by FSIs as the essential missing piece for understanding the $\Omega^-\to \Xi^-\pi^0$ decay, providing a unified framework for the $\Omega^-$ hadronic weak decays. 

In brief, we find that a consistent description of all three decay channels emerges only when the following elements are combined: (i) The large $\Lambda K^-$ rate is dominated by one type of pole term since there is no cancellation. (ii) The $\Xi^0\pi^-$ rate is well-described by the direct, color-allowed weak transition. (iii) The observed $\Xi^-\pi^0$ rate is reproduced through significant constructive interference between the color suppressed tree amplitude and the FSI amplitudes fed from the $\Xi^0\pi^-$ and $\Lambda K^-$ channels. Our results provide a unified picture of $\Omega^-$ hadronic weak decays, demonstrating the indispensable role of non-perturbative long-distance dynamics. The weak decays of $\Omega^-$ provide important information about the interplay between the strong and weak interactions that is complementary to the weak decays of the ground-state spin-1/2 octet baryons.

\begin{acknowledgments}

We would like to thank Dr. Hui-Hua Zhong for useful discussions on solving the numerical wave function of baryons. This work is partly supported by the National Key R\&D Program of China under Grant No. 2023YFA1606703, and by the National Natural Science Foundation of China under Grant Nos. 12575094, 12435007, 12361141819, 12235018 and 12265009. This work is also supported by the Gansu Province Postdoctor Foundation.

\end{acknowledgments}

\appendix
\begin{appendix}

\section{The wave function of hadrons}\label{app:spin and flavor wf}

The flavor and spin wave function can be constructed respectively through the SU(3) and SU(2) symmetries. For the three quark systems, the direct product decomposition of the flavor SU(3) symmetry is $\boldsymbol{3}\otimes\boldsymbol{3}\otimes\boldsymbol{3}=\boldsymbol{10}^s\oplus\boldsymbol{8}^{\rho}\oplus\boldsymbol{8}^{\lambda}\oplus\boldsymbol{1}^a$. Therefore, for the ground-state baryons, there will be a completely symmetric decuplet with $J^P=3/2^+$ and a mixed-symmetry octet with $J^P=1/2^+$. Combining together the spin and flavor symmetry, the spin-flavor wave functions follow an SU(6) symmetry and usually denoted by $|N_6,^{2\bm{S}+1}N_3\rangle$ in the literature, where $N_6$ and $N_3$ represent the dimensions of the SU(6) and SU(3) representations, respectively, and $\bm{S}$ stands for the quantum number of the total spin of a baryon state. The spatial wave functions satisfy the $O(3)$ symmetry under a rotation transformation. For a baryon system containing three quarks, the spatial wave function $\psi_{NLL_z}^{\sigma}=[\psi_{n_{\rho l_{\rho}m_{\rho}}}(\boldsymbol{p}_{\rho})\otimes\psi_{n_{\lambda l_{\lambda}m_{\lambda}}}(\boldsymbol{p}_{\lambda})]_{NLL_z}$ is composed of $\rho$- and $\lambda$- mode spatial wave functions. The quantum numbers $n_{\rho}$, $l_{\rho}$ and $m_{\rho}$ [or $n_{\lambda}$, $l_{\lambda}$, $m_{\lambda}$] stand for that for the radial excitation, relative orbital angular momentum and its $z$ component for the $\rho$-mode [or $\lambda$-mode] wave function, respectively. While $N$, $L$ and $L_z$ stand for the principal quantum number, the quantum numbers of total orbital angular momentum and its $z$ component, respectively. They are defined by $N=2(n_{\rho}+n_{\lambda})+l_{\rho}+l_{\lambda}$, $|l_{\rho}-l_{\lambda}|\leq L\leq l_{\rho}+l_{\lambda}$ and $L_z=m_{\rho}+m_{\lambda}$. Furthermore, the superscript $\sigma$ appearing in wave functions represents their permutation symmetries.

The total wave functions of a baryon can be expressed as $|qqq\rangle_A=|color\rangle_A\otimes|spin,flavor,spatial\rangle_S$. Based on the requirement of the SU(6)$\otimes$O(3) symmetry, one can obtain the configurations for the spin-flavor-spatial part. The total wave function with quantum number $\bm{J}^P=3/2^{+}$ and $\bm{J}^P=1/2^{\pm}$ denoted by $|N_6,^{2\bm{S}+1}N_3,N,\bm{L},\bm{J}^P\rangle$ can be constructed as follows:   
\begin{eqnarray}
&&|56,^410,0,0,\frac32^+\rangle=\phi^s\chi^s\psi_{000}^s,\\
      &&|56,^28,0,0,\frac12^+\rangle=\frac{1}{\sqrt{2}}(\phi^{\rho}\chi^{\rho}+\phi^{\lambda}\chi^{\lambda})\psi_{000}^s,\\
      &&|70,^28,1,1,\frac12^-\rangle=\sum_{L_z+S_z=J_z}\langle 1L_z;\frac12 S_z|\frac12 J_z\rangle \times \nonumber \\
      && \!\!\! \frac{1}{{2}}\left[(\phi^{\rho}\chi^{\lambda}+\phi^{\lambda}\chi^{\rho})\psi_{11L_z}^{\rho}+(\phi^{\rho}\chi^{\rho}-\phi^{\lambda}\chi^{\lambda})\psi_{11L_z}^{\lambda}\right],\\
      &&|70,^28,1,1,\frac32^-\rangle=\sum_{L_z+S_z=J_z}\langle 1L_z;\frac12 S_z|\frac32 J_z\rangle \times \nonumber \\
      && \!\!\! \frac{1}{{2}}\left[(\phi^{\rho}\chi^{\lambda}+\phi^{\lambda}\chi^{\rho})\psi_{11L_z}^{\rho}+(\phi^{\rho}\chi^{\rho}-\phi^{\lambda}\chi^{\lambda})\psi_{11L_z}^{\lambda}\right],\\
      &&|70,^48,1,1,\frac12^-\rangle=\sum_{L_z+S_z=J_z}\langle 1L_z;\frac32 S_z|\frac12 J_z\rangle \times \nonumber \\
      &&\frac{1}{\sqrt{2}}\left[\phi^{\rho}\chi^s\psi_{11L_z}^{\rho}+\phi^{\lambda}\chi^s\psi_{11L_z}^{\lambda}\right].
\end{eqnarray}

The wave function of pseudoscalar mesons is
\begin{eqnarray}
\Phi_{000}(\boldsymbol{p}_1,\boldsymbol{p}_2) &=& \delta^3(\boldsymbol{p}_1+\boldsymbol{p}_2-\boldsymbol{P})\phi_{P}\chi_{0,0}^a \nonumber \\
&& \times  \psi_{000}^s(\boldsymbol{p}_1,\boldsymbol{p}_2),
\end{eqnarray}
where $\chi_{0,0}^s$ is the wave function:
\begin{align}
    \begin{split}
        \chi_{0,0}^a=\frac{1}{\sqrt{2}}(\uparrow\downarrow-\downarrow\uparrow),
    \end{split}
\end{align}
and $\phi_{P}$ is the flavor wave function and we adopt the de Swart's convention~\cite{deSwart:1963pdg, Lu:2024ajt}:
{
\begin{align}
    \vec{\pi}&=\begin{pmatrix}
        -\pi^+\\ \pi^0\\ \pi^-
    \end{pmatrix}=\begin{pmatrix}
        u\bar{d}\\ \frac{1}{\sqrt{2}}(-u\bar{u}+d\bar{d})\\ -d\bar{u}
    \end{pmatrix},\\
    K&=\begin{pmatrix}
        K^+\\ K^0
    \end{pmatrix}=\begin{pmatrix}
        u\bar{s}\\ d\bar{s}
    \end{pmatrix},\bar{K}=\begin{pmatrix}
        \bar{K}^0\\ -K^-
    \end{pmatrix}=\begin{pmatrix}
        s\bar{d}\\ -s\bar{u}
    \end{pmatrix}.
\end{align}
}

The spatial wave function is expressed as:
\begin{align}
    \begin{split}
        \psi_{000}^s(\boldsymbol{p}_1,\boldsymbol{p}_2)=\frac{1}{\pi^{3/4}R^{3/2}}\text{exp}\left[-\frac{(\boldsymbol{p}_1-\boldsymbol{p}_2)^2}{8R^2}\right],
    \end{split}
\end{align}
where $R$ is the harmonic oscillator strength parameter of mesons.

\section{Transition amplitudes extracted in the quark model}
\label{app:amp}

For convenience, we present below the analytical transition amplitudes obtained using the H.O. wave functions. These expressions are provided for completeness and may be skipped without loss of continuity. The amplitudes are expressed in terms of the baryon polarization quantum numbers. In the following expressions, $\mathcal{M}^{J_f,J_f^z;J_i,J_i^z}$ is abbreviated as $\mathcal{M}^{J_f^z;J_i^z}$; $m_q$ denotes the mass of the $u/d$ quarks and $m_s$ the mass of the $s$ quark; $\boldsymbol{k}$ and $\omega_0$ represent the three-vector momentum and energy of the final meson, respectively. To distinguish the H.O. parameters of different hyperons, we label them according to the number of strange quarks in the baryon: the H.O. parameter of $\Omega$ is $\alpha_{sss}$, that of $\Xi$ is $\alpha_{ss\rho}$ and $\alpha_{ss\lambda}$, and that of $\Lambda$ is $\alpha_{s\rho}$ and $\alpha_{s\lambda}$. {Moreover, $R$ is the H.O. parameter of both pion and kaon mesons. }

\begin{widetext}
\begin{itemize}
    \item $\Omega^-\to \Xi^0\pi^-$
\begin{eqnarray}
\mathcal{M}_{\text{DPE,PC}}^{-\frac12;-\frac12} &=& \frac{8G_FV_{ud}V_{us}|\boldsymbol{k}|}{\sqrt{3}\pi^{9/4}}  \Bigg(\frac{R\alpha_{sss}^2\alpha_{ss\lambda}\alpha_{ss\rho}}{\alpha_{sss}^2+\alpha_{ss\rho}^2}\Bigg)^{3/2} \frac{m_q\alpha_{ss\lambda}^2+m_s(\alpha_{sss}^2+2\alpha_{ss\lambda}^2)}{m_q(m_q+2m_s)(\alpha_{sss}^2+\alpha_{ss\lambda}^2)^{5/2}} \nonumber \\
&& \times \text{exp}\Bigg[-\frac{3m_s^2\boldsymbol{k}^2}{(m_q+2m_s)^2(\alpha_{sss}^2+\alpha_{ss\lambda}^2)}\Bigg],\\
\mathcal{M}_{\text{DPE,PV}}^{-\frac12;-\frac12} &=& 0.
\end{eqnarray}

   \item $\Omega^-\to \Xi^-\pi^0$
\begin{eqnarray}
\mathcal{M}_{\text{CS-1,PC}}^{-\frac12;-\frac12} &=& -\frac{4\sqrt6 G_FV_{ud}V_{us}|\boldsymbol{k}|}{9\pi^{9/4}} \Bigg(\frac{\alpha_{sss}^2\alpha_{ss\rho}\alpha_{ss\lambda}R}{\alpha_{sss}^2 + \alpha_{ss\rho}^2}\Bigg)^{3/2}\frac{m_q \alpha_{ss\lambda}^2+m_s(\alpha_{sss}^2+2\alpha_{ss\lambda}^2)}{m_q(m_q+2m_s) (\alpha_{sss}^2 + \alpha_{ss\lambda}^2)^{5/2}} \nonumber \\
&&\times\exp\Bigg[-\frac{3m_s^2\bs k^2}{(m_q+2m_s)(\alpha_{sss}^2 + \alpha_{ss\lambda}^2)}\Bigg],\\
\mathcal{M}_{\text{CS-1,PV}}^{-\frac12;-\frac12} &=& 0.
\end{eqnarray}

   \item $\Omega^-\to \Lambda K^-$
   \begin{eqnarray}
&& \mathcal{M}_{\text{CS-2,PC}}^{-\frac12;-\frac12}= \frac{12\sqrt{6}G_FV_{ud}V_{us}}{\pi^{9/4}}\frac{|\boldsymbol{k}|(R\alpha_{sss}^2\alpha_{s\lambda}\alpha_{s\rho})^{3/2}\big((m_q+m_s)R^2+m_q\alpha_{sss}^2\big)\big((3m_q+m_s)\alpha_{sss}^2+2(2m_q+m_s)\alpha_{s\lambda}^2\big)}{m_q(m_q+m_s) (2m_q+m_s)\big(3\alpha_{sss}^4+4\alpha_{sss}^2\alpha_{s\lambda}^2+6R^2(\alpha_{sss}^2+\alpha_{s\lambda}^2)\big)^{5/2}} \nonumber \\
                && \times \text{exp}\Bigg[-\frac{3m_s^2\boldsymbol{k}^2\Big(6m_qm_s(R^2+\alpha_{sss}^2)+8m_qm_s\alpha_{s\lambda}^2+m_q^2(3R^2+6\alpha_{sss}^2+8\alpha_{s\lambda}^2)+m_s^2\big(3R^2+2(\alpha_{sss}^2+\alpha_{s\lambda}^2)\big)\Big)}{2(m_q+m_s)^2(2m_q+m_s)^2\big(3\alpha_{sss}^4+4\alpha_{sss}^2\alpha_{s\lambda}^2+6R^2(\alpha_{sss}^2+\alpha_{s\lambda})^2\big)}\Bigg],\\
   && \mathcal{M}_{\text{CS-2,PV}}^{-\frac12;-\frac12} = 0, \\
&& \mathcal{M}_{\text{PT,PC}}^{-\frac12;-\frac12}[\Xi^0(\frac12^+)]=  \frac{2\sqrt{6}|\boldsymbol{k}|\big(2m_q(m_q+2m_s)(\alpha_{sss}^2+\alpha_{ss\lambda}^2)-m_q\alpha_{sss}^2\omega_0+(m_q+2m_s)\alpha_{ss\lambda}^2\omega_0\big)(\alpha_{sss}^2\alpha_{ss\lambda}\alpha_{ss\rho})^{3/2}}{3\pi^{3/2}f_K\sqrt{\omega_0}m_q(m_q+2m_s)(\alpha_{sss}^2+\alpha_{ss\lambda}^2)^{5/2}(\alpha_{sss}^2+\alpha_{ss\rho}^2)^{3/2}} \nonumber \\
        && \times\text{exp}\Bigg[-\frac{3m_s^2\boldsymbol{k}^2}{(m_q+2m_s)^2(\alpha_{sss}^2+\alpha_{ss\lambda}^2)}\Bigg] \frac{2m_{\Xi}}{m_{\Lambda}^2-m_{\Xi}^2+im_{\Xi}\Gamma_{\Xi}} \frac{16\sqrt3 G_F V_{ud} V_{us}}{\pi^{3/2}} \Bigg(\frac{\alpha_{s\lambda} \alpha_{s\rho} \alpha_{ss\lambda} \alpha_{ss\rho}}{\alpha_{ss\lambda}^2 + 3\alpha_{ss\rho}^2 + 4\alpha_{s\lambda}^2}\Bigg)^{3/2} \nonumber \\
    && \times  \text{exp}\Bigg[-\frac{3m_s^2(m_q-m_s)^2\boldsymbol{k}^2}{(2m_q+m_s)^2(m_q+2m_s)^2(\alpha_{ss\lambda}^2+3\alpha_{ss\rho}^2+4\alpha_{s\lambda}^2)}\Bigg], \\
&& \mathcal{M}_{\text{PT,PV}}^{-\frac12;-\frac12}[\Xi^0(\frac12^-)]=  \frac{-i4\boldsymbol{k}^2m_s\alpha_{sss}^3\alpha_{ss\rho}^{3/2}\alpha_{ss\lambda}^{5/2}\big(2m_q(m_q+2m_s)(\alpha_{sss}^2+\alpha_{ss\lambda}^2)-m_q\alpha_{sss}^2\omega_0+(m_q+2m_s)\alpha_{ss\lambda}^2\omega_0\big)}{\sqrt{3}\pi^{3/2}f_K\sqrt{\omega_0}m_q(m_q+2m_s)(\alpha_{sss}^2+\alpha_{ss\rho}^2)^{3/2}(\alpha_{sss}^2+\alpha_{ss\lambda}^2)^{7/2}} \nonumber \\
                && \times \text{exp}\Bigg[-\frac{3m_s^2\boldsymbol{k}^2}{(m_q+2m_s)^2(\alpha_{sss}^2+\alpha_{ss\lambda}^2)}\Bigg] \frac{2m_{\Xi}}{m_{\Lambda}^2-m_{\Xi}^2+im_{\Xi}\Gamma_{\Xi}} \frac{i8\sqrt{2}G_FV_{ud}V_{us}}{\pi^{3/2}}\frac{(\alpha_{s\rho}\alpha_{s\lambda}\alpha_{ss\rho}\alpha_{ss\lambda})^{3/2}}{(\alpha_{ss\lambda}^2+3\alpha_{ss\rho}^2+4\alpha_{s\lambda}^2)^{7/2}} \nonumber \\
                && \text{exp}\Big[-\frac{3m_s^2(m_q-m_s)^2\boldsymbol{k}^2}{(2m_q+m_s)^2(m_q+2m_s)^2(\alpha_{ss\lambda}^2+3\alpha_{ss\rho}^2+4\alpha_{s\lambda}^2)}\Big]\times \nonumber \\
                && \bigg(\boldsymbol{k}^2(m_q-m_s)^2m_s^2(\alpha_{ss\lambda}+3\alpha_{ss\rho})\big((m_q+2m_s)\alpha_{ss\lambda}^2-3m_q\alpha_{ss\rho}^2\big)-(2m_q+m_s)^2(m_q+2m_s)^2 \times\nonumber \\
                && (\alpha_{ss\lambda}^2+3\alpha_{ss\rho}^2+4\alpha_{s\lambda}^2)\Big(3(m_q+m_s)\alpha_{ss\lambda}(\alpha_{ss\lambda}-\alpha_{ss\rho})\alpha_{ss\rho}-2\big(2m_s\alpha_{ss\lambda}+m_q(\alpha_{ss\lambda}-3\alpha_{ss\rho})\big)\alpha_{s\lambda}^2\Big)\bigg) \nonumber \\
                && /\bigg(m_qm_s(2m_q+m_s)^2(m_q+2m_s)^2\bigg).
        \end{eqnarray}
   
\end{itemize}
\end{widetext}

\end{appendix}

\bibliography{ref.bib}

\end{document}